# Highlights

## Dynamics in a Low-Rank Separable Field Cellular Automaton

Xiaorui Shi, Mengsha Huang

- A rank-at-most-one separable global field supports extinction, fixed points, cycles, and long transients.
- Survival-birth interval geometry organizes the dynamical phase diagram.
- Long transients form a narrow ridge between two cycle-dominated regimes with distinct dynamical fingerprints.
- Sparse stripe-like cycles reveal how field normalization can lower global density despite wider survival intervals.

# Dynamics in a Low-Rank Separable Field Cellular Automaton

Xiaorui Shi[a,c,*,1], Mengsha Huang[b,*,1]

[a]Weixian College, Tsinghua University, Beijing, China
[b]Department of Physics, Tsinghua University, Beijing, China
[c]Division of Biology and Biological Engineering, California Institute of Technology, Pasadena, CA, USA

ABSTRACT

Cellular automata (CA) are a class of computational models widely used to study the emergence of life-like behavior, typically through local interaction and update rules. Yet globally coupled interaction architectures, which also occur in living and life-inspired systems, are less commonly represented in traditional CA. Here, we introduce a Separable-Field Cellular Automaton (SFCA), in which conventional local-neighborhood interactions are replaced by updates governed by a normalized global coupling field. Each cell is updated according to the normalized field $q = n/n_{\max}$, with survival and birth controlled by two threshold intervals, $\mathcal{S}$ and $\mathcal{B}$. Systematic scans over interval widths and positions revealed four outcome classes: extinction, fixed points, cycles, and long transients. The resulting phase diagram was organized by the relative geometry of the survival and birth intervals: fixed points were enriched when $\mathcal{B}$ was contained within $\mathcal{S}$, whereas long transients concentrated near the boundary between partial overlap and no overlap. A fine scan across this transition revealed a narrow but persistent long-transient ridge separating two qualitatively distinct cycle-dominated regimes: a dense, high-change-rate branch approximating global period-2 alternation and a sparse, low-change-rate, stripe-like branch. Damage-spreading analyses were consistent with a basin-competition interpretation in which prolonged transients reflect delayed selection between these cyclic attractor families, while providing no evidence for sustained exponential trajectory separation. Finite-size analyses further showed that the long-transient ridge persists across all tested lattice sizes, although its position shifts systematically with system size. These results show that structured long-lived dynamics can arise under separable global coupling with a field matrix of rank at most one, suggesting that nontrivial collective organization can persist even under strongly compressed interaction architectures.

## 1. Introduction

A central question in artificial life is whether complex collective organization requires complex interactions. Cellular automata (CA), a class of computational models, have long provided a minimal framework for addressing this question, in which discrete components governed by simple update rules can generate rich spatiotemporal dynamics. Classical examples such as Conway's Game of Life demonstrate that local Moore-neighborhood interactions can produce fixed structures, oscillations, propagating patterns and prolonged transients (Gardner (1970), Wolfram (1984), Crutchfield and Kaneko (1988)). More recent quantitative analyses of Life-like cellular automata have further shown that substantial dynamical complexity can be maintained with relatively parsimonious local rule sets, reinforcing the role of simple cellular rules as a useful substrate for studying emergent organization (Peña and Sayama (2021)).

Cellular automata also remain an active substrate for contemporary Artificial Life research rather than merely a historical model. Recent work has renewed attention to self-reproduction and evolution in spatially distributed computational media (Sayama and Nehaniv (2025)). Recent studies have explored related themes in distinct cellular-automaton frameworks, including emergent hierarchical self-replication in a binary two-dimensional cellular automaton (Yang (2025)) and evolutionary dynamics in a mass-conserving continuous cellular automaton (Plantec, Hamon, Etcheverry, Chan, Oudeyer and Moulin-Frier (2025)). Although these studies pursue different goals, they share a common emphasis on how rich global organization can arise within spatially distributed dynamical media, typically through local or finite-range interactions.

One related development has been the use of trainable cellular automata. Growing neural cellular automata have demonstrated that shared local update rules can be optimized to generate and regenerate prescribed global patterns

*Co-corresponding authors: Xiaorui Shi and Mengsha Huang.
xs326@cantab.ac.uk (X. Shi); huangms22@outlook.com (M. Huang)
ORCID(s): 0009-0009-0378-8116 (X. Shi)
[1]Xiaorui Shi and Mengsha Huang contributed equally to this work and share first authorship.

(Mordvintsev, Randazzo, Niklasson and Levin (2020)). More recent work has extended neural cellular automata to high-resolution outputs and diverse spatial domains while retaining their local self-organizing architecture. Pajouheshgar et al. (Pajouheshgar, Xu, Abbasi, Mordvintsev, Jakob and Süsstrunk (2026)) also identify strictly local information propagation as one factor limiting NCA scaling, as it impedes long-range cell communication. Together with recent developments in binary and continuous cellular automata, these studies demonstrate the remarkable capacity of local or finite-range interactions to generate organized global behavior, while also revealing, in some contexts, limitations associated with strictly local information propagation and motivating the exploration of complementary interaction architectures.

Nevertheless, the complexity of living and life-inspired systems is not restricted to locally interacting architectures. Globally coupled, computation-like dynamics can also arise when individual components contribute to a shared system-level variable or field that subsequently feeds back onto those components. For instance, engineered protein dimerization networks can perform decision-making computations in which collective outputs are encoded by the concentrations and mutual affinities of interacting molecular species (Parres-Gold, Levine, Emert, Stuart and Elowitz (2025)). At another scale, synaptic normalization imposes a population-level constraint that rescales synaptic strengths and can influence learning dynamics in neural systems (Mayzel and Schneidman (2024)).

From the cellular-automaton perspective, nonlocal coupling has also been explored beyond the conventional fixed-neighborhood setting. Early studies explored such extensions through nonlocal interaction structures or update rules that incorporate global context (Li (1992); Culik (1995)), while more recent work has demonstrated that power-law long-range coupling in probabilistic cellular automata can qualitatively reorganize their dynamical phase structure (Pizzi, Nunnenkamp and Knolle (2021)). In the infinite-range limit of the latter, the site-dependent local densities become identical across the lattice, reducing the spatial dynamics to a deterministic zero-dimensional map for the global density. These approaches demonstrate that relaxing locality can qualitatively alter cellular-automaton dynamics, but they do not address the particular case in which global information is retained as a spatially structured yet strongly constrained low-rank field. This motivates a more specific question: what forms of collective dynamics can arise under such a compressed global interaction architecture?

Against this background, we introduce the Separable-Field Cellular Automaton (SFCA) to investigate whether long-lived dynamics can arise under a globally coupled yet strongly constrained interaction architecture, rather than from conventional local Moore-neighborhood interactions. In the SFCA, the global field is deliberately restricted to a low-rank form, imposing a stringent constraint on the effective degrees of freedom of the interaction field. Our coupling rule is inspired by Kauffman's autocatalytic-set perspective, which posits that collective organization can emerge from the joint action of distributed catalytic relations rather than from a central controller (Kauffman (1986), Farmer, Kauffman and Packard (1986)). Binding interactions involving distinct substrate moieties can contribute substantially to catalysis by positioning substrates and favoring catalytically productive configurations (Narlikar, Gopalakrishnan, McConnell, Usman and Herschlag (1995); Schwans, Kraut and Herschlag (2009)). Motivated by this component-wise view of molecular recognition, we abstract the interaction into two coarse-grained compatibility factors and, as a deliberate simplifying assumption, impose separability and represent their joint effect by a product. Taking the row and column activity profiles as these two factors yields an instantaneous field matrix given by the outer product of two vectors and therefore of rank at most one. The SFCA is not intended as a literal model of autocatalytic chemistry, but as a minimal dynamical abstraction of multi-factor compatibility under rank-one coupling.

We ask how the geometry of the survival and birth intervals organizes the long-term dynamics of this system. Systematic scans reveal extinction, fixed points, cycles, and long transients in distinct regions of parameter space. Fixed points are enriched when the birth interval is contained within the survival interval, whereas long transients form a narrow ridge near the boundary between partial overlap and no overlap. This ridge separates a dense, rapidly alternating cycle regime from a sparse, stripe-like, low-change-rate cycle regime, consistent with delayed selection between two cyclic attractor families. The SFCA therefore provides a minimal model in which rank-one field coupling generates prolonged dynamics organized by survival–birth interval geometry.

## 2. Model and methods

### 2.1. Separable-field update rule

The system is a binary cellular automaton on a periodic $H \times W$ lattice (Fig. 1). The state at generation $t$ is denoted by $X_t(j,i) \in \{0,1\}$, where 1 indicates an alive cell. Instead of counting the Moore neighborhood of each cell, the model first computes the column and row occupancies

$$c_i(t) = \sum_{j=1}^{H} X_t(j,i), \qquad r_j(t) = \sum_{i=1}^{W} X_t(j,i). \tag{1}$$

These one-dimensional sums are locally blurred using nearest-neighbor periodic wrapping,

$$C_i(t) = c_{i-1}(t) + c_i(t) + c_{i+1}(t), \qquad R_j(t) = r_{j-1}(t) + r_j(t) + r_{j+1}(t), \tag{2}$$

and the unnormalized field at cell $(j,i)$ is defined as

$$n_t(j,i) = R_j(t)C_i(t). \tag{3}$$

The field is then normalized by its lattice maximum,

$$q_t(j,i) = \frac{n_t(j,i)}{n_{\max}(t)}, \qquad n_{\max}(t) = \max_{j,i} n_t(j,i), \tag{4}$$

with the convention that an all-dead lattice remains all dead.

Two closed intervals determine the synchronous update rule. An alive cell survives if $q_t(j,i) \in \mathcal{S} = [S_{\text{low}}, S_{\text{high}}]$, while a dead cell is born if $q_t(j,i) \in \mathcal{B} = [B_{\text{low}}, B_{\text{high}}]$. Thus

$$X_{t+1}(j,i) = \begin{cases} 1, & X_t(j,i) = 1 \text{ and } q_t(j,i) \in \mathcal{S}, \\ 1, & X_t(j,i) = 0 \text{ and } q_t(j,i) \in \mathcal{B}, \\ 0, & \text{otherwise.} \end{cases} \tag{5}$$

The normalization by $n_{\max}$ is essential: the same absolute configuration can become more or less permissive depending on how strongly the current pattern concentrates the field.

## 2.2. Parameterization of survival and birth intervals

For the coarse scans, endpoints were placed on an 18-level grid. The survival and birth widths were

$$w_S = S_{\text{high}} - S_{\text{low}}, \qquad w_B = B_{\text{high}} - B_{\text{low}}. \tag{6}$$

In the width-only scan, every pair $(w_S, w_B)$ was averaged over all legal interval positions satisfying $S_{\text{low}} < B_{\text{low}}$. In geometry-focused scans, we fixed the offset

$$\Delta_{\text{low}} = B_{\text{low}} - S_{\text{low}} \tag{7}$$

and used the two boundaries $B_{\text{high}} = S_{\text{high}}$ and $B_{\text{low}} = S_{\text{high}}$ to divide the plane into three interval-geometric regions: $\mathcal{B} \subseteq \mathcal{S}$, partial overlap, and no overlap.

For the fine transition scan, endpoints were placed on a 180-level grid. Unless otherwise stated, we used $S_{\text{low}} = 10/180 = 1/18$ and $\mathcal{B} = [60/180, 160/180] = [6/18, 16/18]$, and varied $w_S$ along the transition axis. This axis crosses the long-transient ridge observed in the coarse phase diagram while keeping the birth interval fixed.

## 2.3. Simulation protocol and outcome classification

Initial states were independent Bernoulli random lattices with density $\rho_0 = 0.25$, except in the initial-density control experiment where $\rho_0$ was varied from 0.05 to 0.95. Most scans used a $75 \times 100$ lattice. Coarse scans were run for 2000 generations. Fine transition and survival analyses were run for up to 100,000 generations. A full-history hash buffer was used to detect repeated states during each run.

Each trajectory was assigned to one of four outcome classes. Extinction was recorded when all cells were dead. A fixed point was recorded when the state repeated with period 1. A cycle was recorded when a previous state repeated with period at least 2. A long transient was recorded when the maximum generation did not detect an extinction, fixed point, or cycle. The long-transient class should therefore be interpreted as “not observed to terminate within the specified observation window”, not as proof of non-periodicity for infinite time.

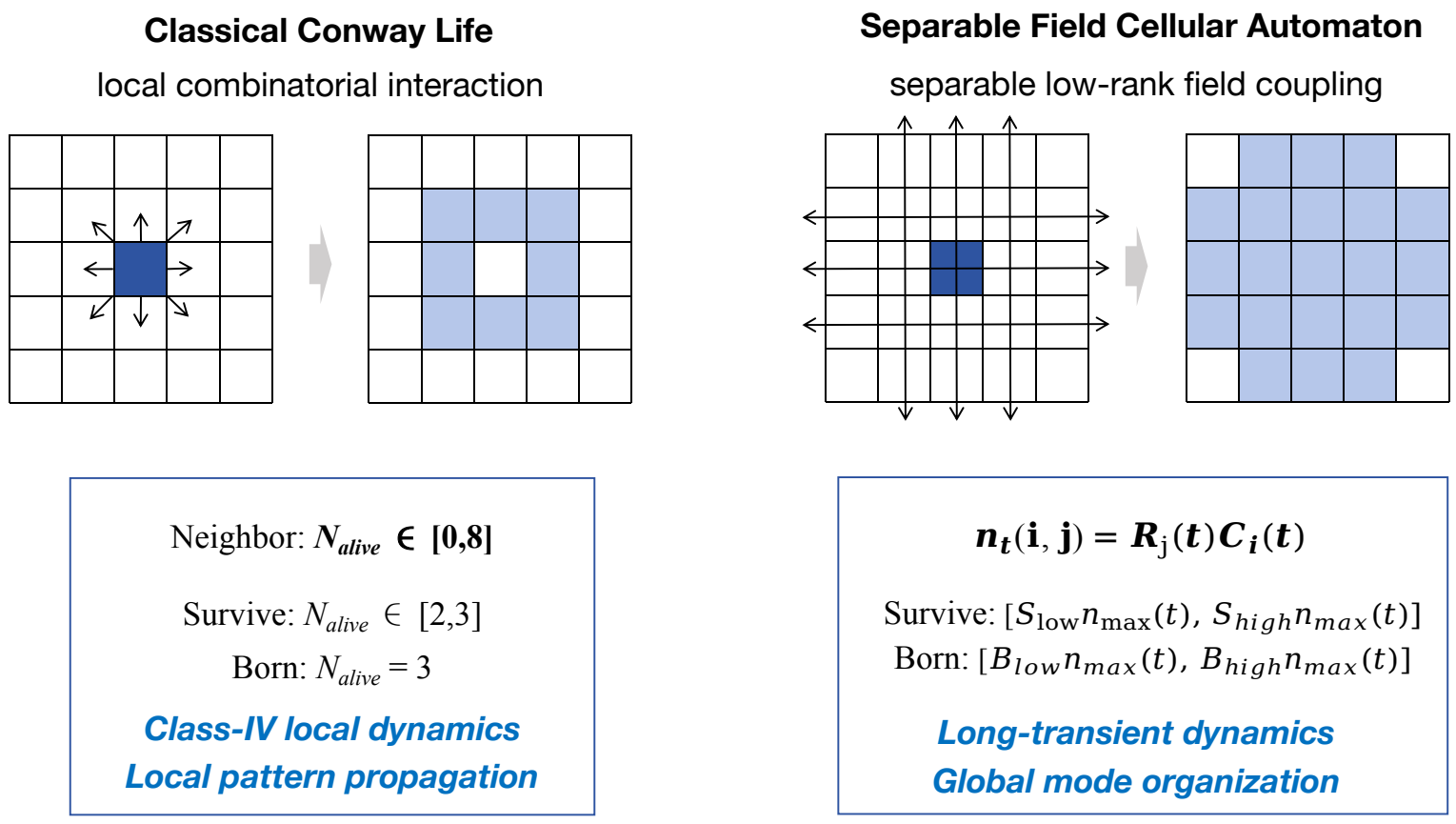

**Figure 1:** Comparison between Classical Conway Life and the Separable Field Cellular Automaton (SFCA). Conway Life is a classical local cellular automaton in which each cell updates based on the number of alive neighbors in its 3×3 neighborhood. Cells survive if they have 2–3 alive neighbors and are born with exactly 3 neighbors. This produces Class-IV local dynamics with localized pattern propagation, including gliders, oscillators, and still lifes. The Separable Field Cellular Automaton (SFCA) updates each cell according to a rank-1 separable row-column interaction field, $n_t(i,j) = R_j(t)C_i(t)$. Survival and birth are determined relative to the normalized field ($[S_{low}n_{max}(t), S_{high}n_{max}(t)]$ for survival; $[B_{low}n_{max}(t), B_{high}n_{max}(t)]$ for birth). The SFCA exhibits long-transient dynamics and global mode organization.

## 2.4. Dynamical measurements

To characterize attractor dynamics, we measured density, change rate, stripe score, and damage spreading. The density is

$$\rho(t) = \frac{1}{HW}\sum_{j,i} X_t(j,i), \tag{8}$$

and the change rate is

$$\chi(t) = \frac{1}{HW}\sum_{j,i} \left|X_{t+1}(j,i) - X_t(j,i)\right|. \tag{9}$$

For cycle-class runs, we aggregated the mean density and mean change rate within the detected cycle or over the recorded trajectory after convergence.

To quantify row- or column-aligned spatial heterogeneity, we define a simple stripe score. For a lattice $X$, let $\bar{X}_j$ denote the mean occupancy of row $j$ and $\bar{X}_i$ the mean occupancy of column $i$. We define

$$\sigma_{\text{stripe}}(X) = \text{Var}_j(\bar{X}_j) + \text{Var}_i(\bar{X}_i). \tag{10}$$

Uniform or spatially homogeneous configurations have low stripe scores, whereas configurations organized into horizontal or vertical bands produce larger row- or column-level variance and therefore higher stripe scores.

To characterize sensitivity to local perturbations, we used the standard damage-spreading approach, in which two trajectories initialized from nearly identical configurations are evolved under the same update rule and their separation is monitored over time (Bagnoli, Rechtman and Ruffo (1992)). In our implementation, the two initial states differed by a single cell, and their separation was quantified using the normalized Hamming distance

$$d_H(t) = \frac{1}{HW}\sum_{j,i} |X_t(j,i) - Y_t(j,i)|. \tag{11}$$

We further computed a finite-time effective perturbation-growth rate,

$$\lambda_{\text{eff}}(t) = \frac{1}{t} log(\frac{\tilde{d}_H(t)}{\tilde{d}_H(0)}) \tag{12}$$

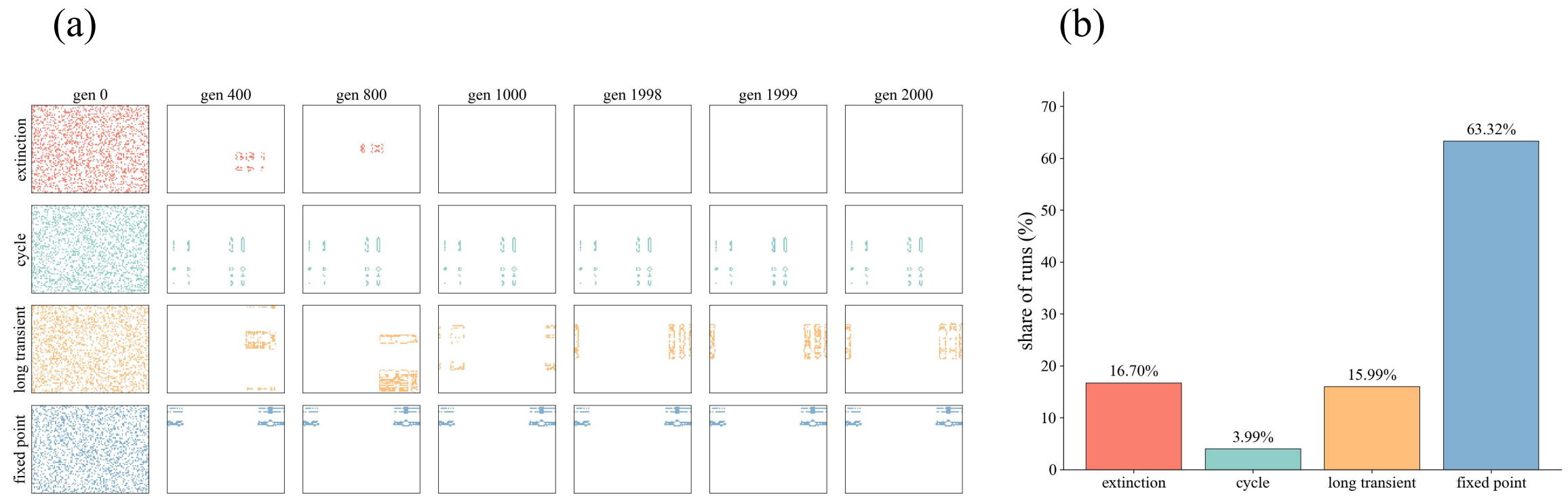

**Figure 2:** Representative outcome classes under a single SFCA rule. (a) Example trajectories leading to extinction, fixed point, cycle, and long transient under $\mathcal{S} = [3/18, 11/18]$ and $\mathcal{B} = [7/18, 9/18]$. Each row shows snapshots from an independent random initialization at generations 0, 400, 800, 1000, 1998, 1999, and 2000. Colored cells indicate alive cells and blank cells indicate dead cells. The extinction example reaches the all-dead state by generation 825; the fixed-point example reaches a static state by generation 275; the cycle example enters a period-2 orbit by generation 132; and the long-transient example remains unclassified within the 2000-generation observation window. (b) Outcome composition from 100,000 independent random initializations under the same rule. The same parameter setting can generate extinction, fixed points, cycles, and long transients depending on the initial condition.

This quantity measures the average logarithmic rate of perturbation amplification between the initial condition and generation $t$, rather than an asymptotic Lyapunov exponent.

Phase-shift and translation controls were used to test whether large distances could be explained by the same cycle observed at different phases or by spatially shifted versions of similar patterns.

# 3. Results

## 3.1. A single SFCA rule can support all four outcome classes

Before scanning the full parameter space, we first verified that the SFCA can support multiple long-term outcomes under a single rule. For the representative rule $\mathcal{S} = [3/18, 11/18]$ and $\mathcal{B} = [7/18, 9/18]$, 100,000 independent random initial states yielded all four outcome classes within a 2000-generation window: 16.70% extinction, 3.99% cycle, 15.99% long transient, and 63.32% fixed point (Fig. 2b). Representative snapshots show that these outcomes are visually distinct despite sharing the same rule parameters (Fig. 2a).

## 3.2. A global parameter scan reveals four dynamical regimes

We next scanned the phase diagram using $w_S$ and $w_B$ as the two axes, with interval endpoints discretized in units of $1/18$. For each width pair, all legal interval positions satisfying $S_{\text{low}} < B_{\text{low}}$ were sampled, and 1000 trajectories for each width combination were simulated on a $75 \times 100$ grid for 2000 generations. When certain width combinations include more than one group of $S_{\text{low}}, S_{\text{high}}, B_{\text{low}}, B_{\text{high}}$ combinations, the trajectories were evenly distributed among each group. The resulting four-class outcome heatmaps showed a strongly organized phase diagram rather than a monotonic response to interval width. Extinction was concentrated near the lower-left corner, where both survival and birth intervals were narrow. Fixed points dominated when the survival interval was wide and the birth interval was narrow. Cycle-rich regions occupied broad bands outside this fixed-point domain, while long transients appeared primarily as intermediate transition regions (Fig. 3a).

The outcome maxima emphasize this organization. In the width-only scan, extinction reached 88.3% at the smallest survival and birth widths. Cycle share reached 100% in a wide-birth, narrow-survival region. Long-transient share reached 57.0% in the interior transition band. Fixed-point share reached 95.0% in a wide-survival, narrow-birth region. These observations suggest that the balance between survival permissiveness and birth permissiveness is not sufficient by itself; the relative placement of $\mathcal{S}$ and $\mathcal{B}$ must also be considered.

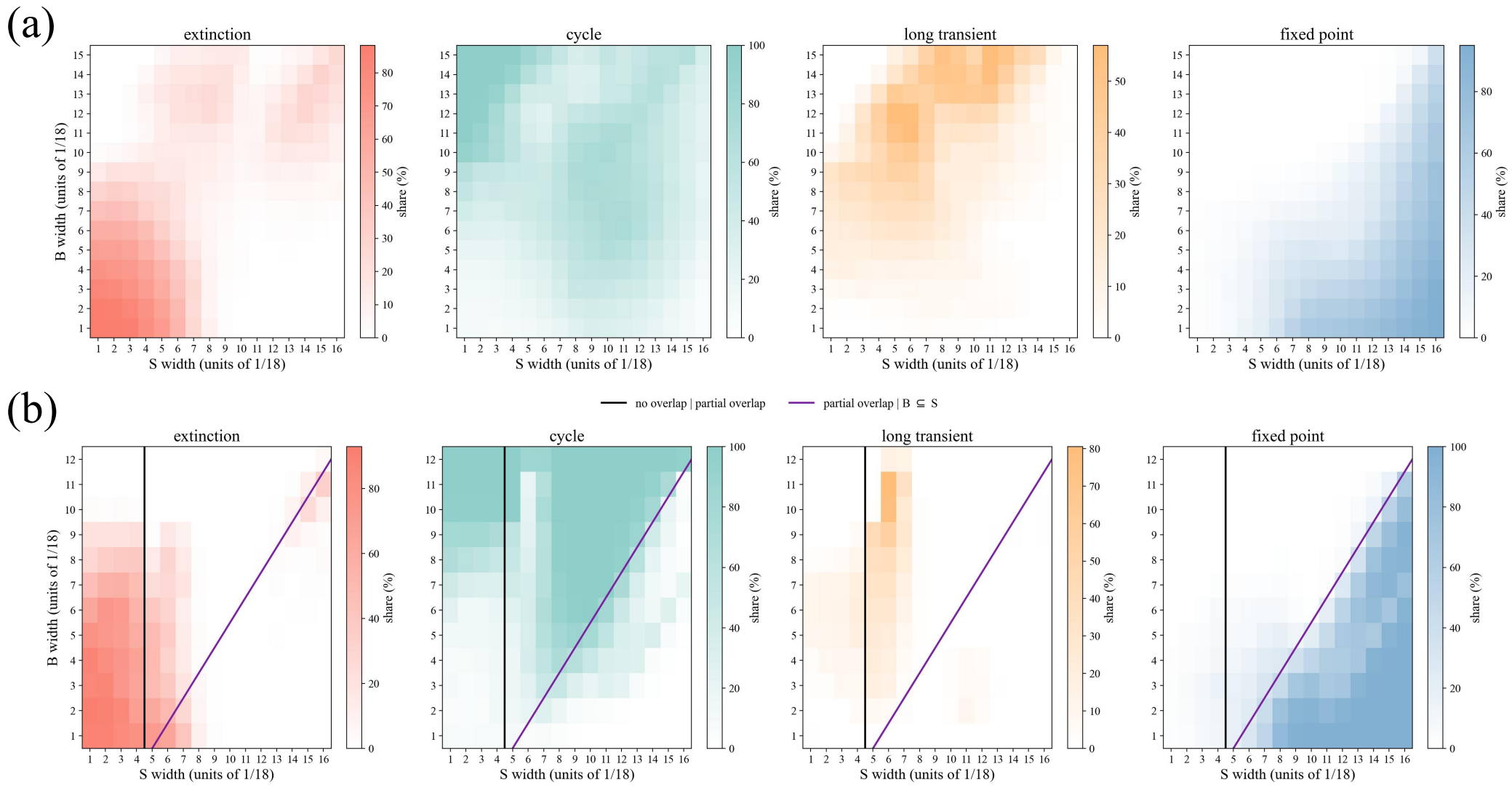

**Figure 3:** (a)Global outcome phase diagram over survival width $w_S$ and birth width $w_B$. Each cell averages over legal interval positions and 1000 random initial states. The four panels show the fraction of extinction, cycle, long-transient, and fixed-point outcomes. The scan reveals extinction near narrow intervals, fixed points in wide-survival/narrow-birth regions, and long-transient transition bands that separate cycle-rich regions. (b)Interval geometry for the offset $\Delta_{\text{low}} = 5/18$. The boundary $B_{\text{low}} = S_{\text{high}}$ separates no-overlap from partial-overlap regimes, while $B_{\text{high}} = S_{\text{high}}$ separates partial overlap from $\mathcal{B} \subseteq \mathcal{S}$. Fixed points are enriched where the birth interval is contained within the survival interval, whereas long transients form a transition ridge near the overlap boundary.

## 3.3. Survival-birth interval geometry organizes the phase diagram

To isolate interval geometry from width alone, we fixed $\Delta_{\text{low}} = B_{\text{low}} - S_{\text{low}}$. For $\Delta_{\text{low}} = 5/18$, the boundaries $B_{\text{low}} = S_{\text{high}}$ and $B_{\text{high}} = S_{\text{high}}$ divide the width plane into no-overlap, partial-overlap, and $\mathcal{B} \subseteq \mathcal{S}$ regions. This geometric decomposition captured two major features of the phase diagram (Fig. 3b). First, the $\mathcal{B} \subseteq \mathcal{S}$ region was enriched for fixed points, consistent with the idea that cells born under $\mathcal{B}$ are then likely to satisfy the survival condition. Second, long transients were concentrated near the boundary where the upper edge of the survival interval approaches the lower edge of the birth interval.

This geometry initially explains why the long-transient region separates two cycle-rich domains. When $S_{\text{high}}$ is below $B_{\text{low}}$, survival and birth are driven by disjoint field ranges. When $S_{\text{high}}$ crosses $B_{\text{low}}$, a persistence band appears in which a cell can satisfy both survival and birth conditions depending on its current state. The long-transient ridge lies near this change in rule logic, indicating that delayed convergence is linked to the emergence of competing update mechanisms rather than to random noise in the scan. This organization was also retained when the absolute positions of the survival and birth intervals were shifted while keeping $\Delta_{\text{low}} = 5/18$ fixed (Fig. S5). Similar geometric organization was observed across $\Delta_{\text{low}}$ values from $1/18$ to $8/18$, with the cycle, extinction, fixed-point, and long-transient regions reorganizing systematically relative to the two interval-overlap boundaries (Figs. S1–S4).

## 3.4. Outcome classes are mostly robust to initial density

A possible alternative explanation for the observed phase structure is that it depends strongly on the chosen initial density. To test this, we selected four representative anchor rules enriched for extinction, fixed points, cycles, and long transients, respectively, at the reference density $\rho_0 = 0.25$. The four anchor rules were

$$
\begin{aligned}
\text{extinction:}\quad & \mathcal{S} = [50/180, 100/180], && \mathcal{B} = [80/180, 110/180],\\
\text{fixed point:}\quad & \mathcal{S} = [30/180, 120/180], && \mathcal{B} = [50/180, 70/180],\\
\text{cycle:}\quad & \mathcal{S} = [10/180, 63/180], && \mathcal{B} = [60/180, 160/180],\\
\text{long transient:}\quad & \mathcal{S} = [10/180, 69/180], && \mathcal{B} = [60/180, 160/180].
\end{aligned}
$$

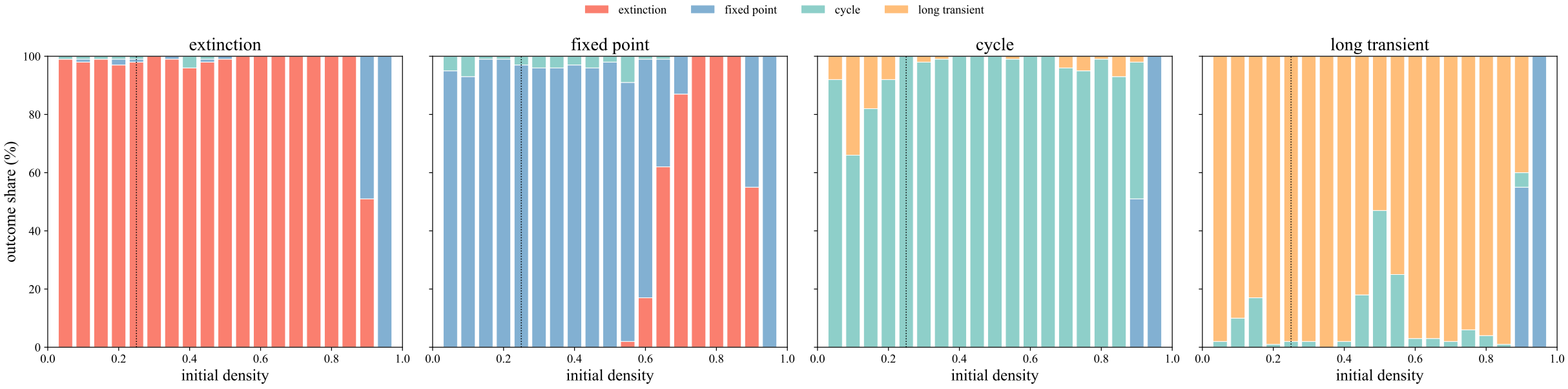

**Figure 4:** Initial-density control. Outcome composition was measured for anchor rules enriched for different outcome classes while varying $\rho_0$ from 0.05 to 0.95. Extinction, cycle, and long-transient anchors show broad robustness around $\rho_0 = 0.25$, whereas fixed-point behavior is more sensitive to high initial density.

For each rule, we varied the initial density from $\rho_0 = 0.05$ to 0.95 in increments of 0.05, with 100 independent random initializations simulated at each density on a $75 \times 100$ lattice for 2000 generations. The extinction, cycle, and long-transient anchors retained their characteristic outcome over broad density ranges around $\rho_0 = 0.25$ (Fig. 4). In particular, the cycle and long-transient anchors remained dominated by their respective outcomes from $\rho_0 = 0.05$ to 0.85. The fixed-point anchor was more sensitive to initial density, with high-density initial conditions increasingly shifting toward extinction.

These results indicate that the principal outcome regimes are not artifacts of a single initial-density choice. Initial density can modulate the accessibility of particular attractors, most notably fixed points, but the cycle- and long-transient-dominated behaviors remain robust over a broad range of initial conditions. We therefore use $\rho_0 = 0.25$ as a representative reference density for the subsequent analyses.

## 3.5. A persistent long-transient peak separates two cycle-dominated regimes

We then focused on the transition axis $S_{low} = 10/180$, $\mathcal{B} = [60/180, 160/180]$, scanning $w_S$ at resolution $1/180$. On this axis, extinction and fixed points were absent; the system showed a direct competition between cycles and long transients. In the 100,000-generation scan over $w_S = 45/180$ to $75/180$, cycle share was 100% on both sides of the transition, whereas long-transient share rose sharply between $57/180$ and $64/180$, reaching 97% at $w_S = 61/180$ (Fig. 5a).

Kaplan-Meier-style survival curves confirmed that the long-transient window remained visible under longer observation times (Fig. 5b, Fig. S6). At $w_S = 50/180, 53/180, 56/180, 65/180$, and $70/180$, all 1000 runs eventually entered detected cycles by 100,000 generations. In contrast, the middle widths retained substantial censored fractions at the endpoint: 62.5% at $58/180$, 94.7% at $60/180$, and 75.0% at $62/180$. Thus, the long-transient peak is not an artifact of the 2000-generation coarse scan; it marks a region with exceptionally delayed attractor entry.

## 3.6. The two cycle regions have distinct dynamical fingerprints

Although both sides of the long-transient peak were cycle-dominated, their cycles were not dynamically equivalent. On the low-$w_S$ side, cycle-class runs had high density and high change rate. At $w_S = 55/180$, the mean cycle density was $0.470 \pm 0.039$ and the mean change rate was $0.997 \pm 0.007$, indicating almost global alternation between consecutive phases. On the high-$w_S$ side, cycles were sparse and nearly stationary. At $w_S = 70/180$, the mean cycle density was $0.332 \pm 0.045$ and the mean change rate was $0.027 \pm 0.013$ (Fig. 6a). Representative periodic trajectories further illustrate this distinction: a dense period-2 cycle occurs on the low-$w_S$ side, whereas the high-$w_S$ side can exhibit sparse higher-period stripe-like cycles (Fig. S7).

The decrease in cycle density with increasing survival width is initially counterintuitive. A wider survival interval might be expected to preserve more living cells, yet the high-$w_S$ cycle branch was less dense than the low-$w_S$ branch. The stripe score resolved this paradox. Sparse cycles had substantially higher stripe scores than dense cycles: the score was approximately $0.032 \pm 0.010$ at $w_S = 55/180$ but $0.084 \pm 0.020$ at $w_S = 70/180$ (Fig. 6b). The high-$w_S$ branch therefore does not simply maintain fewer cells uniformly; it organizes cells into heterogeneous stripe-like structures.

Because the update rule uses $q = n/n_{max}$, such heterogeneity changes the effective field landscape. Dense bands or hotspots raise $n_{max}$, which lowers the relative field values of cells outside those regions. As a result, high-density regions can remain self-sustaining while low-density regions remain below the birth threshold. Widening the survival interval can therefore stabilize localized structures without increasing the global density. This mechanism distinguishes the sparse, low-change-rate cycle basin from the dense, high-change-rate cycle basin.

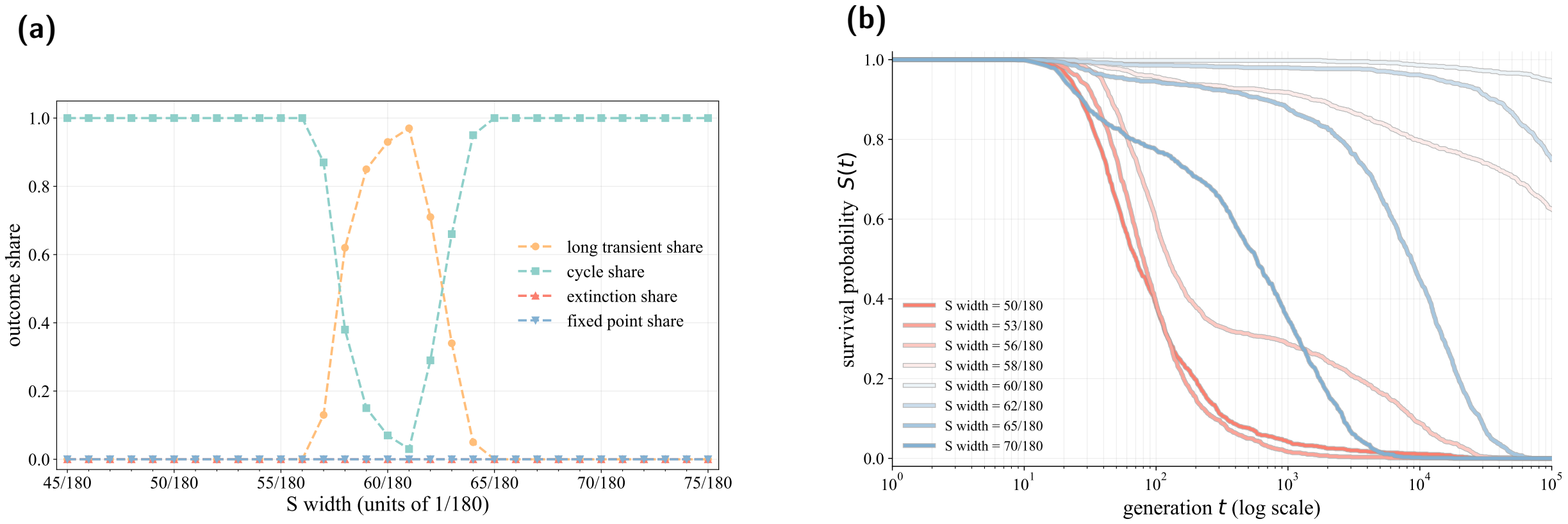

**Figure 5:** Fine transition scan. (a) Outcome shares along the canonical axis $S_{\mathrm{low}} = 10/180$ and $\mathcal{B} = [60/180, 160/180]$. Long transients form a narrow peak between two cycle-dominated plateaus. (b) Kaplan–Meier-style survival curves for selected $w_S$ values, where survival denotes remaining unclassified by extinction, fixed point, or cycle at generation $t$.

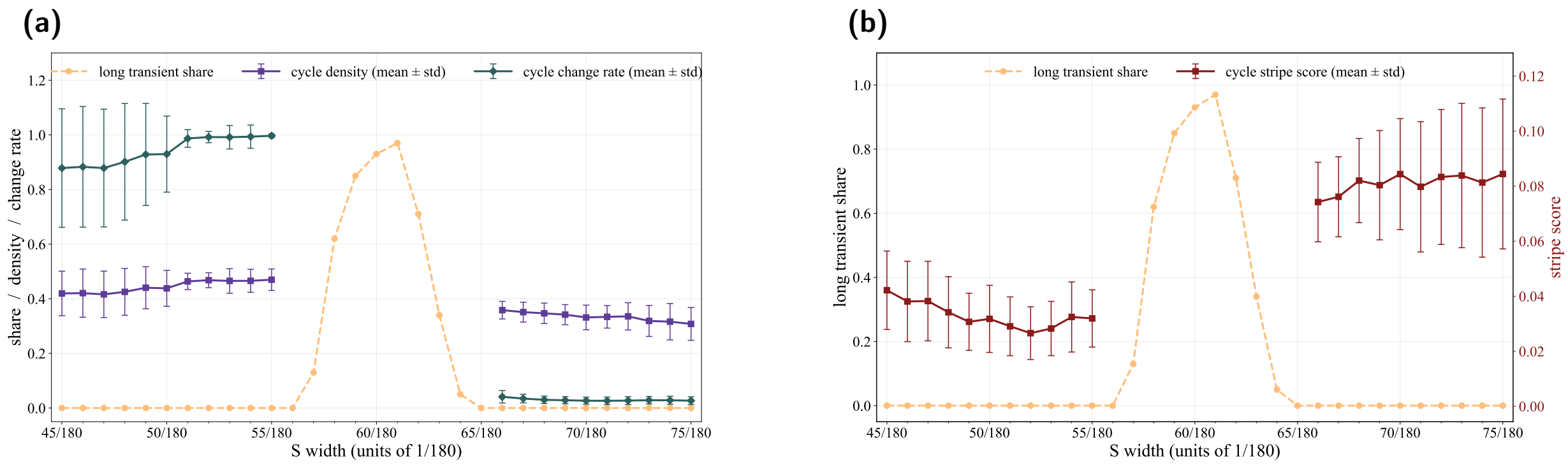

**Figure 6:** Cycle fingerprints on the two sides of the long-transient window. (a) Cycle density and change rate separate a dense, high-change-rate branch from a sparse, low-change-rate branch. (b) Stripe score is higher in the sparse branch, indicating stronger row/column heterogeneity. Error bars show standard deviation across cycle-class runs.

## 3.7. Long transients are consistent with competition between two cycle basins

The combined outcome and fingerprint data support a basin-competition interpretation. The low-$w_S$ cycle branch and high-$w_S$ cycle branch have distinct densities, change rates, and stripe scores, consistent with distinct attractor families rather than simple continuous variations of a single cycle type. The long-transient peak lies between them, precisely where the rule geometry changes from almost disjoint survival and birth ranges to a wider persistence band. We therefore interpret the long-transient ridge as a delayed basin-selection region: trajectories near this boundary do not rapidly commit to either the dense alternating cycle basin or the sparse stripe-like cycle basin.

This interpretation also explains why the long-transient peak is narrow. Far from the boundary, one cycle basin dominates and convergence is rapid or moderately delayed. Near the boundary, neither basin fully stabilizes early, and trajectories can wander for long times through states that retain features of both mechanisms. The long-transient class is therefore not an isolated anomaly but a structured transition zone in rule space.

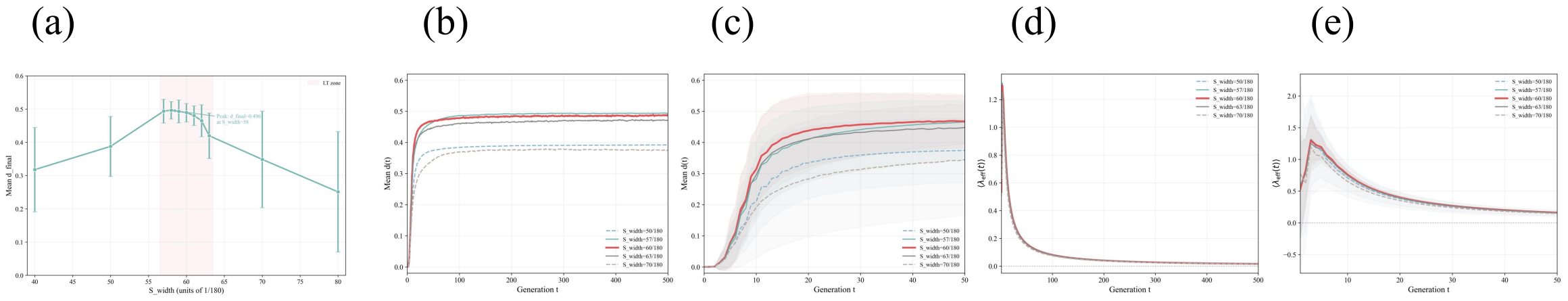

**Figure 7:** Damage spreading across the long-transient transition. (a) Plateau normalized Hamming distance between paired trajectories initialized from states differing by a single cell, plotted as a function of survival width $w_S$. The shaded region marks the survival-width range in which long transients are enriched. Hamming separation is lower in the two cycle-dominated regions and increases near the long-transient peak. (b) Representative time courses of $d_H(t)$ for selected $w_S$ values. Perturbations spread rapidly from the one-cell difference and approach width-dependent plateaus. (c) First 50 generations of $d_H(t)$ for selected $w_S$ values. (d) Effective finite-time growth rate $\lambda_{\mathrm{eff}}(t)$ computed from changes in $d_H(t)$. The growth rate rises during the early spreading phase but decays toward zero, providing no evidence from this diagnostic for sustained exponential trajectory separation. (e) First 50 generations of the effective finite-time growth rate $\lambda_{\mathrm{eff}}(t)$ for selected $w_S$ values.

## 3.8. Damage spreading is enhanced in the long-transient region

We next characterized sensitivity to perturbations using paired-trajectory damage spreading. The plateau Hamming distance was lower in the two cycle-dominated regions and higher near the long-transient peak (Fig. 7a). The time-dependent distance $d_H(t)$ rose rapidly from the initial perturbation and then approached a width-dependent plateau (Fig. 7b,c). Extended trajectories and higher-resolution early-time traces show the same behavior across a broader set of $w_S$ values (Figs. S8, S9). The corresponding effective growth rate $\lambda_{\mathrm{eff}}(t)$ increased early but decayed toward zero rather than remaining positive indefinitely (Fig. 7d,e). These observations indicate that the long-transient region amplifies perturbations to larger separations than the neighboring cycle basins, but it does not behave as a sustained exponentially divergent chaotic regime.

Two controls were used to test whether the elevated Hamming distance was a measurement artifact (Figure S10-S11). First, we compared phases of the same cycle to determine whether $d_H(t)$ was dominated by phase offsets within a periodic orbit. Second, we tested whether similar patterns shifted up, down, left, or right could account for the observed distance. Both contributions were small. Together, these controls indicate that the observed damage-spreading signal cannot be explained primarily by phase mismatch or global translation.

## 3.9. Finite-size scans support a robust long-transient window

Finally, we examined whether the long-transient window depends on lattice size (Fig. 8a). Fine scans along the canonical transition axis were repeated on grids of size $50\times37$, $100\times75$, $150\times112$, and $200\times150$. The corresponding outcome compositions and survival curves across representative $w_S$ values are shown in Fig. S12. The long-transient peak was present at all tested sizes, reaching approximately 87% on the smallest grid within the 100,000-generation observation window. For grids of size $100 \times 75$ and larger, the peak long-transient fraction approached saturation at approximately 97% or higher. The position of the peak nevertheless shifted systematically toward smaller $w_S$ as lattice size increased, indicating a finite-size dependence in the location, but not the existence, of the long-transient window.

To examine whether this finite-size shift was accompanied by changes in the organization of the normalized interaction field, we measured the spatial standard deviation of $q = n/n_{\max}$ at each generation during the first 2000 generations (Fig. 8b). Additional field statistics, including the coefficients of variation of the row, column, and full-field profiles and the ratio $n_{\max}/\langle n\rangle$, showed corresponding grid-size-dependent organization across representative survival widths (Fig. S13). We compared three representative survival widths, $w_S = 55/180$, $60/180$, and $62/180$, across the same four grid sizes. The grid-size dependence of the normalized-field fluctuations qualitatively tracked the corresponding differences in long-transient enrichment across these parameter settings. This correspondence suggests that the finite-size dependence of the long-transient window is associated with changes in normalized-field organization, although the present analysis does not establish a causal relationship.

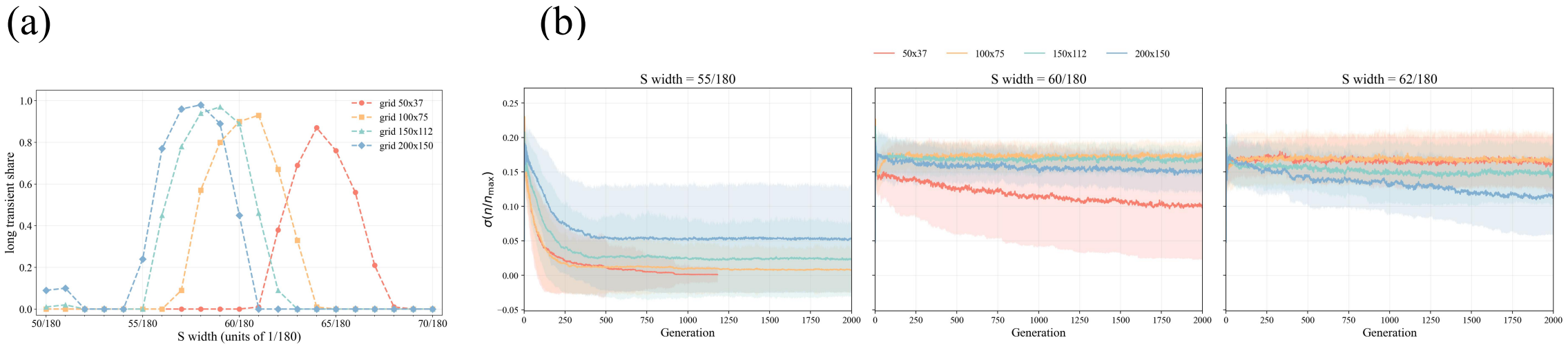

**Figure 8:** Finite-size behavior of the long-transient window. (a) Long-transient share across the canonical transition axis for grids of size $50 \times 37$, $100 \times 75$, $150 \times 112$, and $200 \times 150$. The long-transient peak is present at all tested sizes, increases from the smallest grid to larger grids, and shifts toward smaller $w_S$ as lattice size increases. (b) Spatial standard deviation of the normalized field $q = n/n_{\max}$ during the first 2000 generations for selected survival widths. Across grid sizes, higher normalized-field fluctuations are associated with stronger long-transient enrichment.

## 4. Discussion

The main result is that long transients in the SFCA are geometrically organized. They arise near specific relationships between the survival and birth intervals, especially near the boundary where $S_{\text{high}}$ approaches $B_{\text{low}}$. Long-transient behavior itself is not unique to the SFCA: nonlocal cellular automata have previously been shown to exhibit long transient times (Li (1992)), and power-law long-range coupling in probabilistic cellular automata can qualitatively reorganize dynamical phase structure (Pizzi et al. (2021)). The distinctive feature of the SFCA is that long transients are not observed merely as isolated trajectories or isolated rules, but form a narrow and reproducible ridge in parameter space whose location is tied to the relative geometry of the survival and birth intervals. Thus, under rank-one separable coupling, prolonged dynamics are systematically organized by a low-dimensional rule geometry.

The interval-containment result provides a simple explanation for the fixed-point-rich region. When $\mathcal{B} \subseteq \mathcal{S}$, the field values that allow dead cells to become alive also allow alive cells to remain alive. Once a viable pattern is formed, the rule therefore contains a built-in stabilizing tendency. This does not guarantee convergence to a fixed point from every initial condition, but it explains why fixed points dominate in that geometric region of the phase diagram.

The long-transient ridge is more subtle. Along the canonical transition axis, the ridge separates two cycle branches with sharply different fingerprints. The low-$w_S$ branch behaves like a dense alternating attractor, with nearly all cells flipping each generation, whereas the high-$w_S$ branch behaves like a sparse structured attractor in which only a small fraction of cells change. Because the long-transient peak lies between these two branches, the results are consistent with a basin-competition interpretation in which trajectories undergo delayed selection between qualitatively distinct cyclic attractor families. This interpretation is further supported by the damage-spreading results: nearby trajectories separate more strongly near the ridge than inside either cycle-dominated regime, but the finite-time effective perturbation-growth rate decays toward zero rather than remaining persistently positive.

In related distributed dynamical systems, interactions between random Boolean networks have been shown to shift ordered and chaotic dynamics toward critical behavior (Braccini, Baldini and Roli (2025)). The SFCA results, however, caution against identifying a long transient lifetime with critical or chaotic dynamics by itself. Perturbations in the long-transient region are rapidly amplified to finite macroscopic separations, but the observed saturation of Hamming distance and decay of the effective growth rate do not provide evidence for sustained exponential divergence. Within the measurements considered here, the coincidence between the long-transient ridge and the transition between two distinct cyclic fingerprints is therefore more consistent with delayed attractor selection than with a sustained chaotic regime.

Long-lived behavior in nonlocal discrete systems can also arise through mechanisms distinct from either sustained chaos or attractor competition. Lazarides and Pizzi showed that weak long-range interactions can produce apparent bistability in a system that formally possesses only one stable phase, because nucleation of sufficiently large destabilizing droplets becomes extremely unlikely (Lazarides and Pizzi (2026)). The SFCA differs mechanistically from this scenario: its dynamics are deterministic once the initial configuration is specified, and the long-transient ridge does not rely on stochastic nucleation or defect-growth barriers. Instead, it appears at the boundary between two cycle-dominated dynamical regimes. This comparison illustrates that very long observed lifetimes in nonlocal systems can arise from qualitatively different mechanisms and should not, by themselves, be taken as evidence for a single notion of metastability or criticality.

The density paradox highlights another distinctive consequence of the SFCA coupling architecture: rank-one global coupling does not imply a spatially uniform mean field. In the infinite-range limit of the long-range probabilistic cellular automaton studied by Pizzi et al., site-dependent local densities become identical and the spatial system reduces to a deterministic zero-dimensional map for a single global density (Pizzi et al. (2021)). In the SFCA, by contrast, the field remains spatially structured because it is generated from the outer product of row and column activity profiles. The subsequent normalization by $n_{\max}$ makes this spatial structure dynamically important: stripe-like configurations can concentrate the field and increase $n_{\max}$, thereby reducing normalized field values elsewhere on the lattice. The system can therefore support a lower global cell density even when the survival interval is wider. This behavior emphasizes that the normalization converts spatial concentration of the low-rank field into a global relative constraint on cell updates.

Several limitations remain. First, the long-transient class is operationally defined by the finite observation window. Because the SFCA on a finite lattice is a deterministic finite-state system, every trajectory must eventually enter a periodic orbit; trajectories classified here as long transients are therefore those for which no repeated state was detected within the observation window. Longer simulations or analytical state-space arguments would be required to distinguish exceptionally long transients from eventual cycles reached only after much longer times or having periods beyond the current observation window. Second, the basin-competition interpretation is supported by the location of the long-transient ridge, the contrasting cycle fingerprints, and the damage-spreading analysis, but a direct mapping of attractor basins would be needed to establish this mechanism more firmly. Third, the finite-size scans demonstrate persistence of the long-transient window across the tested lattice sizes, but the systematic shift of the peak position shows that an asymptotic limit has not yet been established. In particular, the present data do not justify identifying the ridge with a thermodynamic critical point or assigning it to a universality class.

Future work should therefore focus on direct basin mapping, targeted perturbation experiments, and systematic finite-size scaling of the position, width, and lifetime distribution of the long-transient ridge. The separable structure of the field also suggests the possibility of approximate lower-dimensional descriptions based on row and column activity profiles. Such reductions could clarify how the two cyclic regimes arise, why the transition between them is sharp, and how stripe-like attractors become stabilized as the survival–birth geometry changes. More broadly, recent Artificial Life studies have demonstrated rich self-organizing and evolutionary dynamics in both binary and continuous cellular automata (Yang (2025); Plantec et al. (2025)). The SFCA addresses a complementary question: rather than enriching the local rule or state space, it asks how much dynamical organization can persist when the interaction architecture itself is strongly compressed. The present results suggest that coupling structure, field normalization, and effective interaction rank provide additional axes along which minimal Artificial Life models can be organized and compared.

## 5. Conclusion

We introduced a normalized separable-field cellular automaton in which binary cells interact through a globally coupled field constrained to rank at most one. Despite this strongly compressed interaction architecture, the SFCA exhibits extinction, fixed points, cycles, and long transients across structured regions of parameter space. The geometry of the survival and birth intervals provides a simple organizing principle for these outcomes: fixed points are enriched when the birth interval is contained within the survival interval, whereas long transients concentrate near the boundary between partial overlap and no overlap.

Along a fine transition scan, the long-transient ridge separates two qualitatively distinct cycle-dominated regimes: a dense, high-change-rate branch and a sparse, stripe-like, low-change-rate branch. Cycle fingerprints and damage-spreading analyses are consistent with a basin-competition interpretation in which prolonged transients reflect delayed selection between these cyclic attractor families. At the same time, perturbations saturate rather than exhibiting sustained exponential divergence, indicating that the long lifetime of the ridge should not by itself be interpreted as evidence of sustained chaotic dynamics. Finite-size analysis further shows that the ridge persists across all tested lattice sizes, although its systematic shift with system size leaves its asymptotic scaling an open question.

Together, these results show that severe compression of the interaction field does not eliminate structured collective dynamics. Unlike a scalar global mean field, the rank-one separable field retains spatial organization through its row and column components, while normalization couples that structure back to the entire lattice. The SFCA therefore provides a minimal framework for studying how interaction geometry and global normalization organize long-lived dynamics under low-rank coupling. More broadly, it suggests that the architecture and effective rank of interactions can themselves serve as control dimensions for collective organization, complementary to increasing the complexity of local cellular-automaton rules.

## Acknowledgments

We thank Lulu Qian for helpful discussions on the conceptual development of the model and Yancheng Zhou for proofreading an early version of the manuscript.

## Data and code availability

The simulation code and analysis scripts used in this study are publicly available on GitHub at `https://github.com/huangmengs/SFCA.git`. Generated figures and processed outputs can be reproduced from the repository and the parameter settings described in this manuscript.

## Competing interests

The authors declare no competing interests.

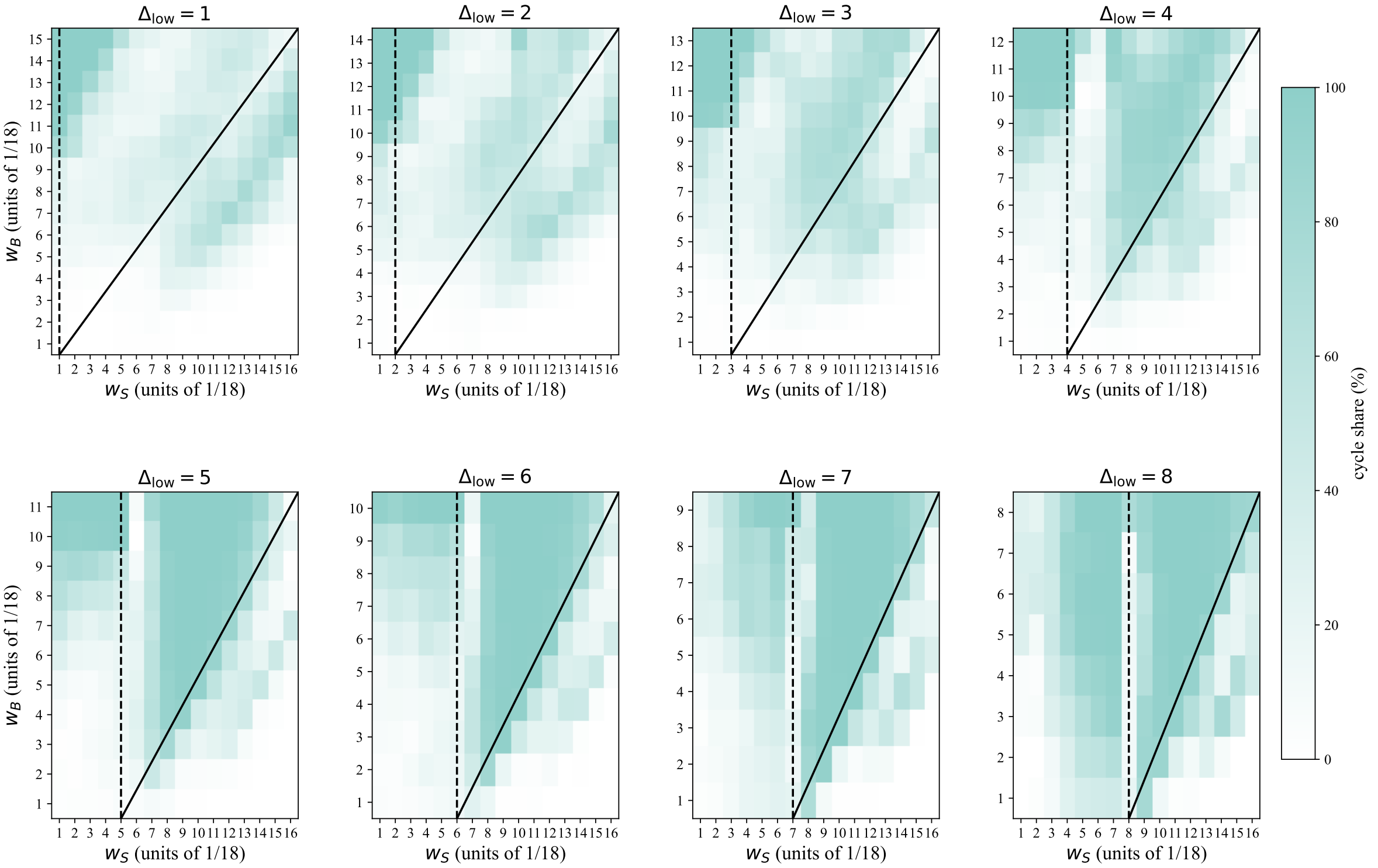

**Figure S1:** Cycle phase maps on the $w_S \times w_B$ plane for $\Delta_{\text{low}} = B_{\text{low}} - S_{\text{low}}$ ranging from 1/18 to 8/18. The black dashed line denotes $B_{\text{low}} = S_{\text{high}}$, and the black solid line denotes $B_{\text{high}} = S_{\text{high}}$. Parameters: $S_{low} = 1/18$, $B_{low} = 6/18$, 2000 generations per run, 100 random initial conditions per parameter set, initial density 0.25.

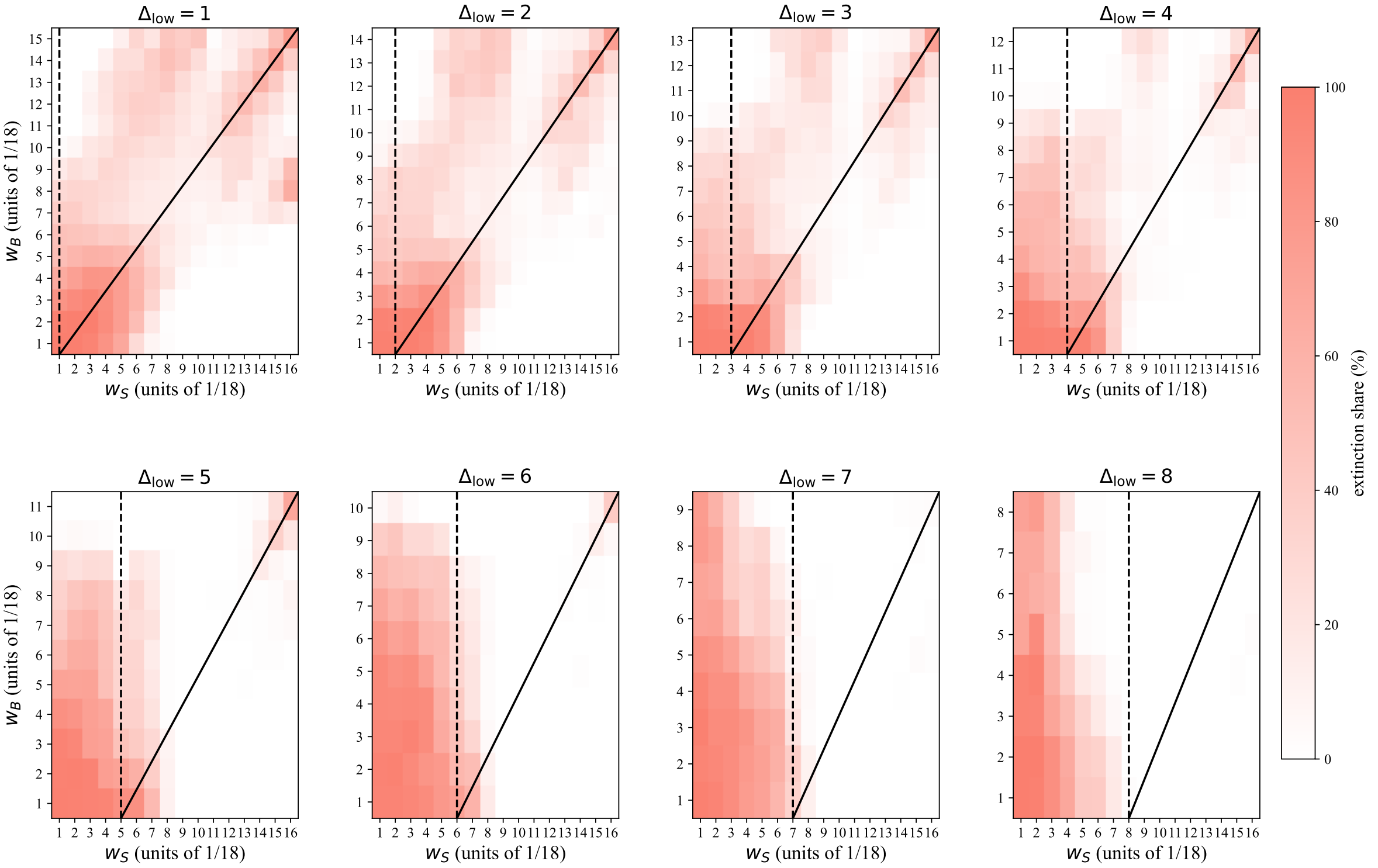

**Figure S2:** Extinction phase map on the $w_S \times w_B$ plane with $\Delta_{low} = B_{low} - S_{low}$ varying from 1/18 to 8/18. Black dashed line: $B_{low} = S_{high}$; black solid line: $B_{high} = S_{high}$. Parameters: $S_{low} = 1/18, B_{low} = 6/18$, 2000 generations per run, 100 random initial conditions per parameter set, initial density 0.25.

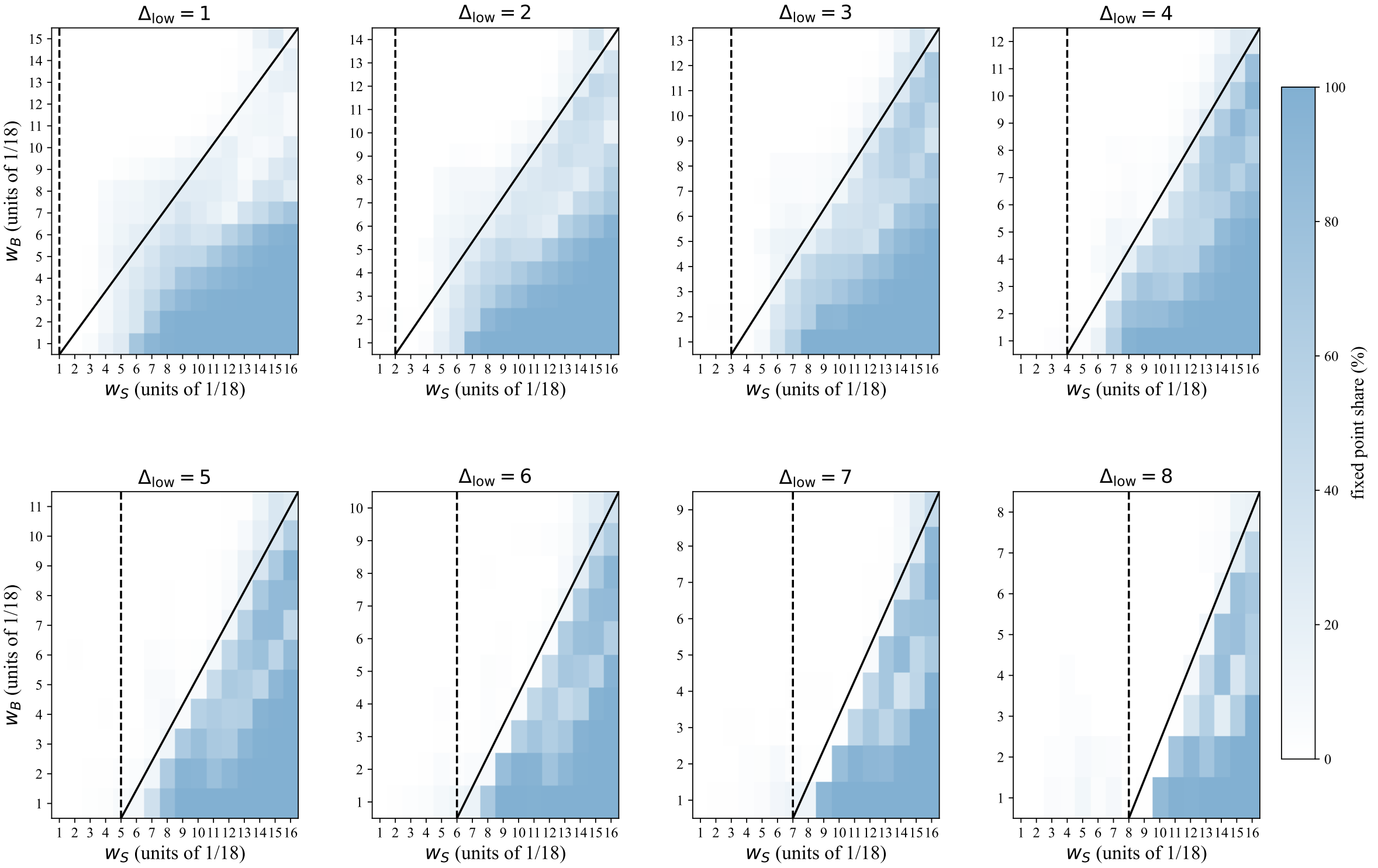

**Figure S3:** Extinction phase map on the $w_S \times w_B$ plane with $\Delta_{\text{low}} = B_{\text{low}} - S_{\text{low}}$ varying from 1/18 to 8/18. Black dashed line: $B_{low} = S_{high}$; black solid line: $B_{high} = S_{high}$. Parameters: $S_{low} = 1/18, B_{low} = 6/18$, 2000 generations per run, 100 random initial conditions per parameter set, initial density 0.25.

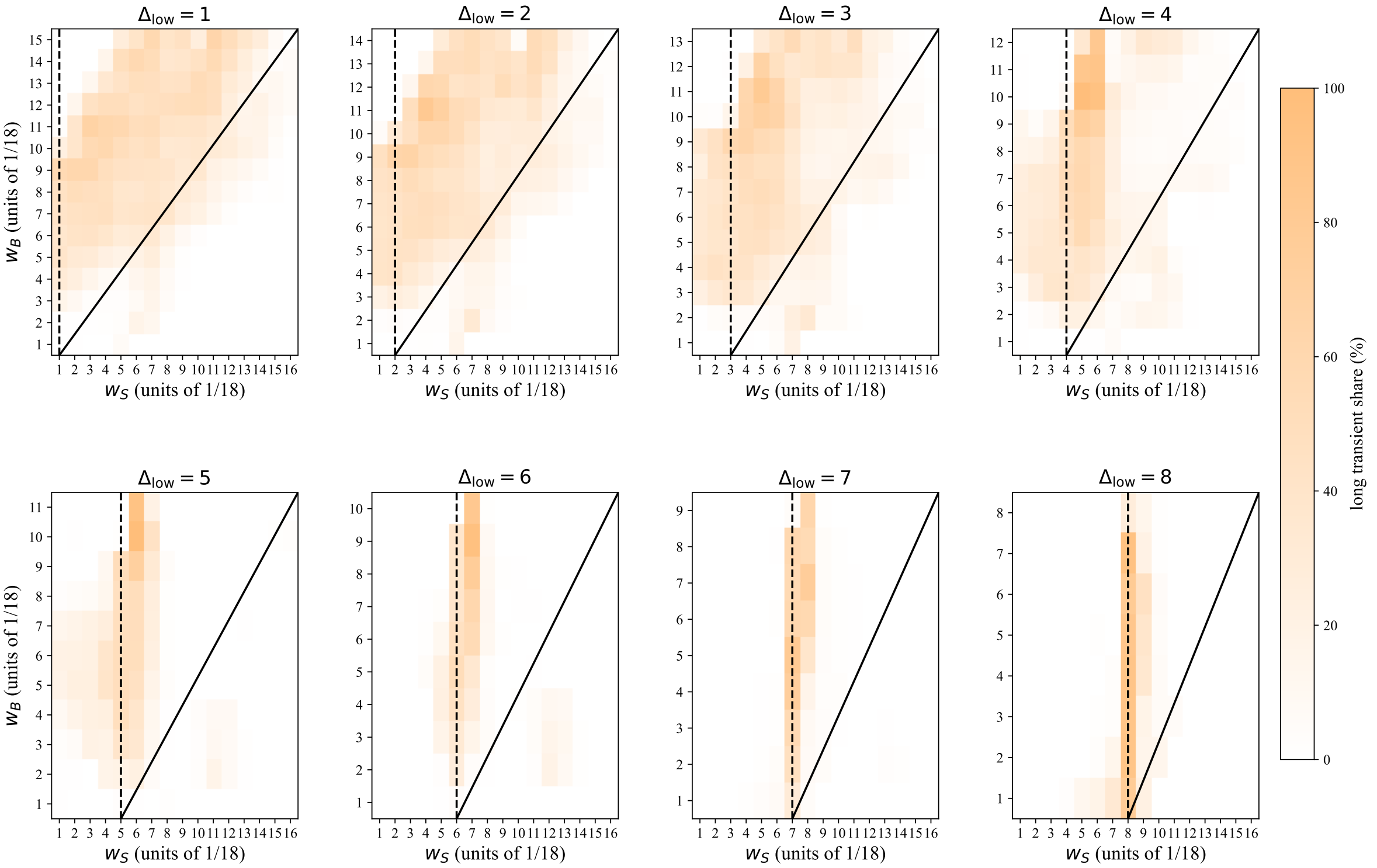

**Figure S4:** Long transient phase map on the $w_S \times w_B$ plane with $\Delta_{low} = B_{low} - S_{low}$ varying from 1/18 to 8/18. Black dashed line: $B_{low} = S_{high}$; black solid line: $B_{high} = S_{high}$. Parameters: $S_{low} = 1/18, B_{low} = 6/18$, 2000 generations per run, 100 random initial conditions per parameter set, initial density 0.25.$B_{low} = S_{high}$

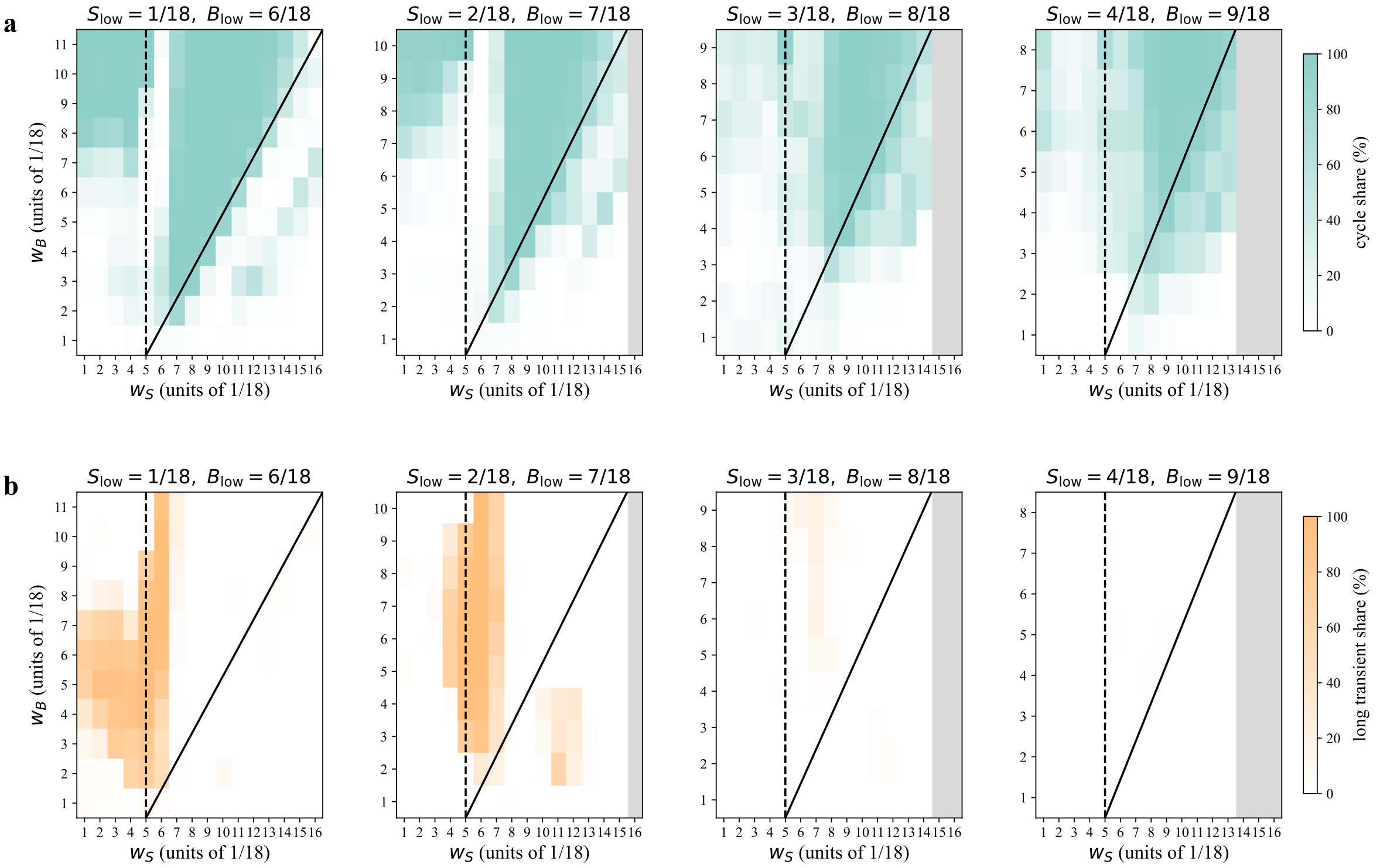

**Figure S5:** (a) Cycle phase map on the $w_S \times w_B$ plane when $\Delta_{low} = B_{low} - S_{low} = 5/18$, and $S_{low}$ vary from 1/18 to 4/18. Black dashed line: $B_{low} = S_{high}$; black solid line: $B_{high} = S_{high}$. Parameters: 2000 generations per run, 100 random initial conditions per parameter set, initial density 0.25. (b) Long transient phase map on the $w_S \times w_B$ plane when $\Delta_{low} = B_{low} - S_{low} = 5/18$, and $S_{low}$ vary from 1/18 to 4/18. White dashed line: $B_{low} = S_{high}$; white solid line: $B_{high} = S_{high}$. Parameters: 2000 generations per run, 100 random initial conditions per parameter set, initial density 0.25.

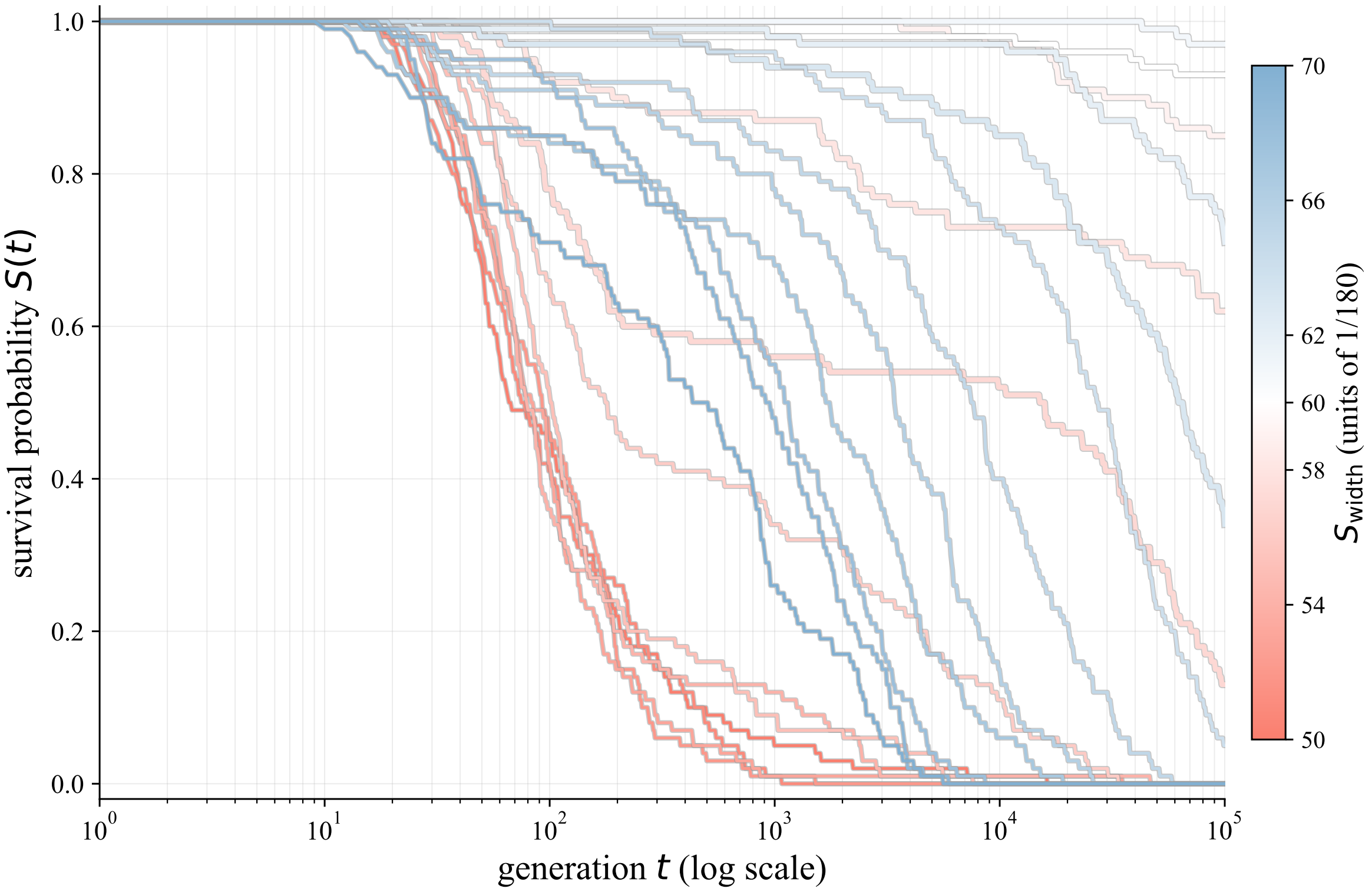

**Figure S6:** Kaplan–Meier-style survival curves of transient lifetime across $w_S$. Survival probability $S(t) = P(\tau > t)$ denotes the fraction of trajectories that have not yet reached extinction, a fixed point, or a detected periodic cycle by generation $t$. Curves are shown for $w_S$ values spanning the transition between the two cycle-dominated regimes. Trajectories classified as long transients at the end of the 100,000-generation observation window are treated as right-censored. Near the long-transient-enriched region, the survival curves remain elevated throughout the observation window, whereas trajectories in the neighboring cycle-dominated regimes terminate substantially earlier. Parameters: $100 \times 75$ grid, $S_{low} = 1/18$, $B_{low} = 6/18$, $B_{high} = 16/18$, initial density 0.25.

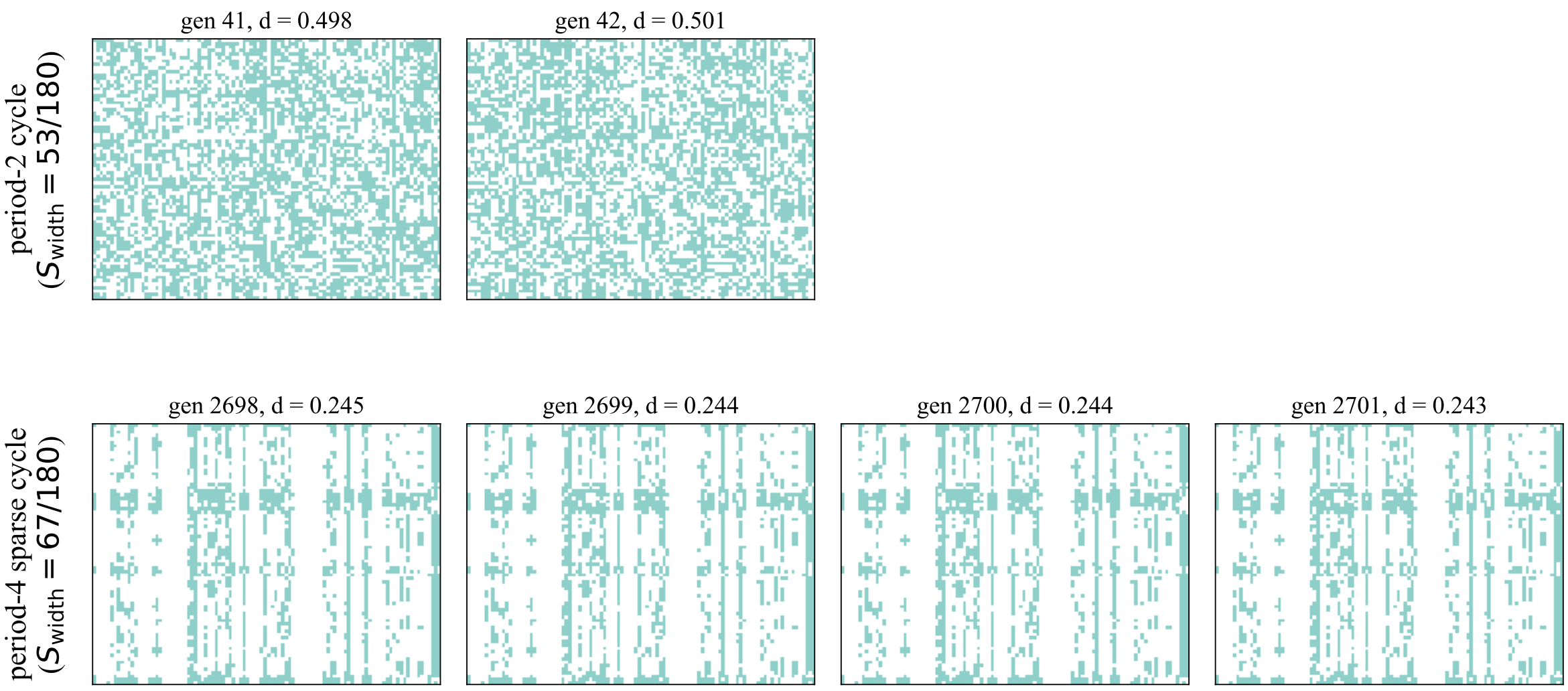

**Figure S7:** Representative dense and sparse cycle trajectories on opposite sides of the long-transient window along the canonical transition axis. The upper row shows a dense period-2 cycle at $w_S = 53/180$, with all two cycle phases shown. The lower row shows a sparse period-4 cycle at $w_S = 67/180$, with all four cycle phases shown. These examples illustrate the qualitatively distinct spatial organizations of the two cycle-dominated regimes separated by the long-transient window.

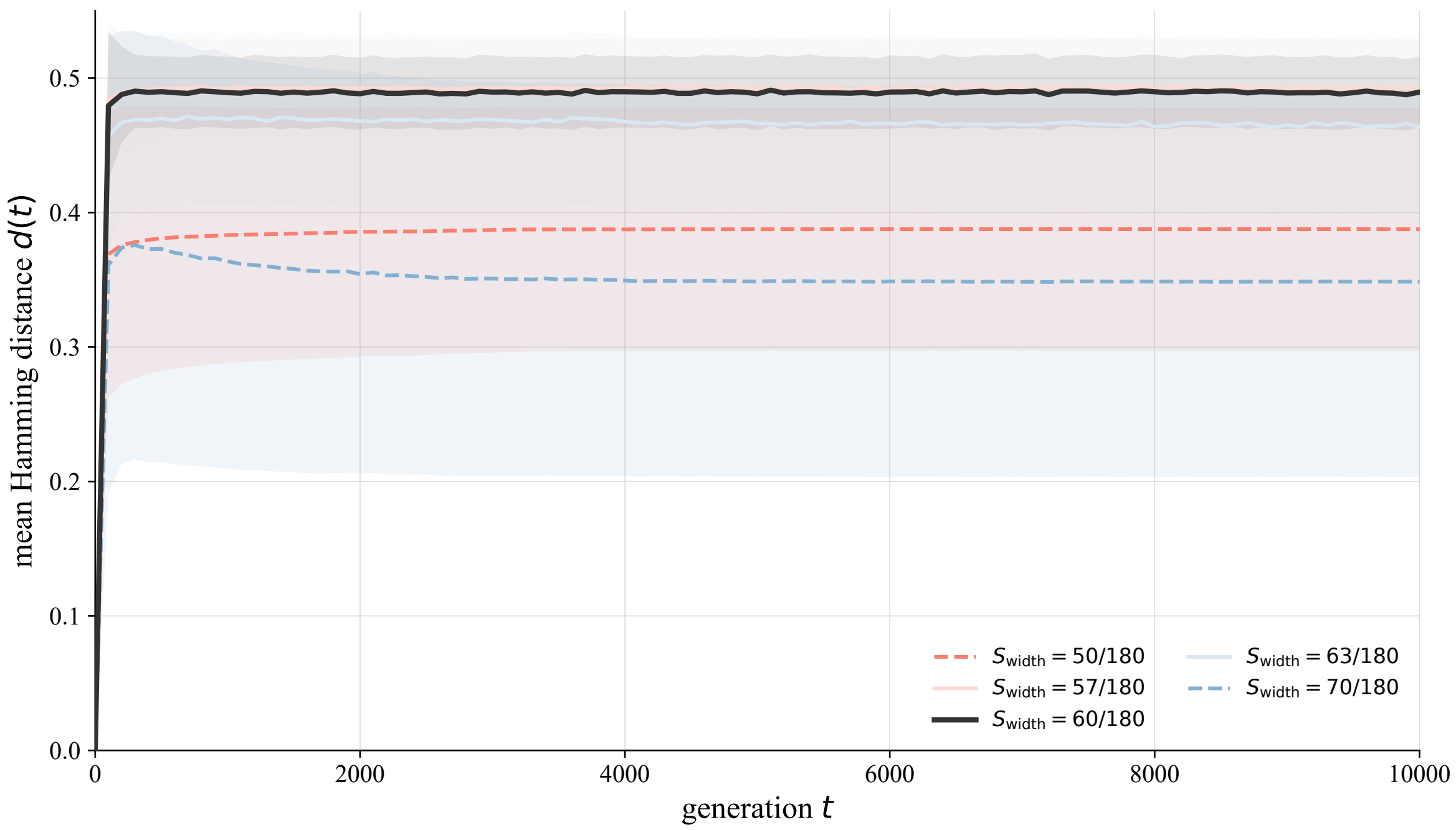

**Figure S8:** Mean normalized Hamming distance $d_H(t)$ between paired trajectories initialized from states differing by a single flipped cell. The Hamming distance between two initially identical $100 \times 75$ grids differing by a single flipped cell is shown as a function of generation t, averaged over 500 independent pairs per $S_{width}$ value. Solid curves correspond to $w_S$ values in the long-transient-enriched region (57/180–63/180), whereas dashed curves show representative values from the neighboring cycle-dominated regimes (50/180 and 70/180). Shaded ribbons indicate $\pm 1$ standard deviation. Parameters: $S_{low} = 1/18$, $B_{low} = 6/18$, $B_{high} = 16/18$, $B_{width} = 10/18$, grid size $100 \times 75$, 100,000 generations per run, initial density 0.25.

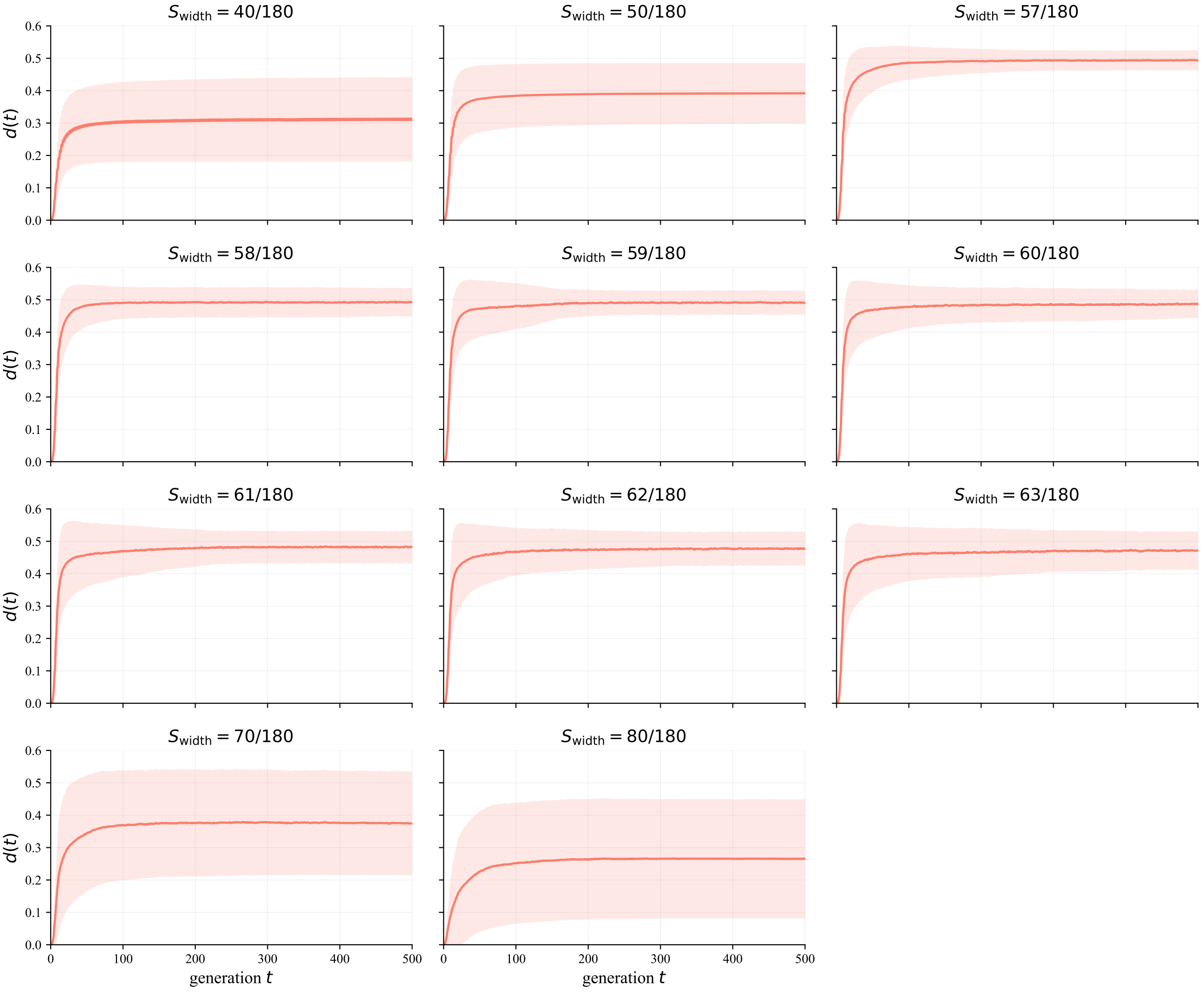

**Figure S9:** High-resolution early-time evolution of the normalized Hamming distance $d_H(t)$ for 11 $w_S$ values. Each panel shows $d_H(t)$ between paired trajectories initialized from states differing by a single flipped cell, recorded every generation for the first 500 generations. Solid curves represent the mean across 500 independent runs; shaded ribbons indicate $\pm 1$ standard deviation. Parameters: $S_{low} = 1/18$, $B_{low} = 6/18$, $B_{high} = 16/18$, $B_{width} = 10/18$, $\Delta_{low} = 5/18$, grid size $100 \times 75$, 500 generations per run, 500 random initial conditions per parameter set, initial density 0.25.

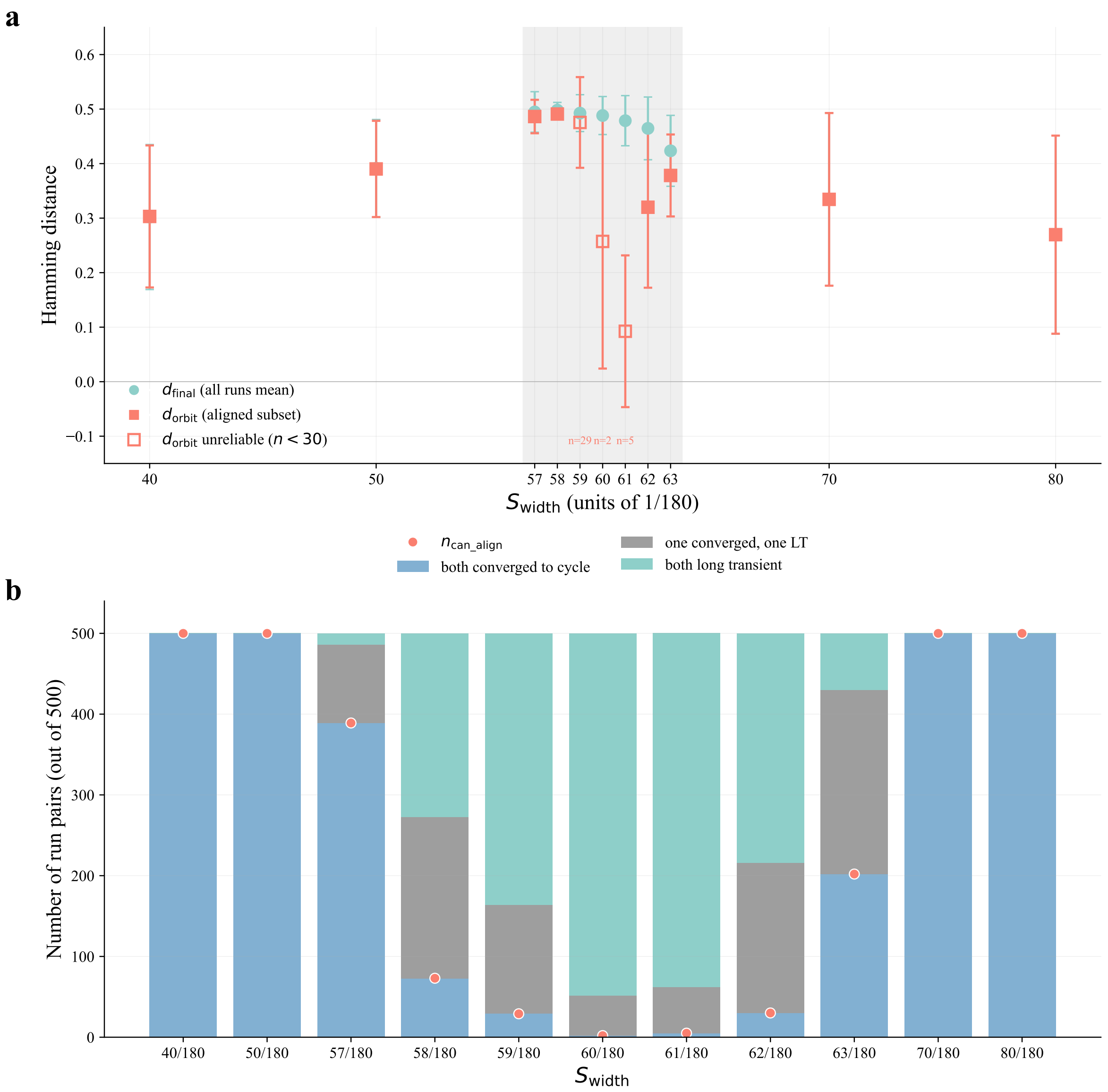

**Figure S10:** Phase-alignment control for damage-spreading measurements. (a) Final normalized Hamming distance, $d_{\text{final}}$, and phase-aligned cycle distance, $d_{\text{orbit}}$, across $w_S$. Open circles show the mean $d_{\text{final}}$ over all 500 paired trajectories at each $w_S$, including both converged and long-transient runs. Solid squares show $d_{\text{orbit}}$, defined as the minimum Hamming distance after phase alignment and computed only for run pairs in which both the reference and perturbed trajectories converged to detectable cycles with cycle length $\leq 10{,}000$ generations. Error bars indicate $\pm 1$ standard deviation. Hollow markers denote $w_S$ values for which fewer than 30 run pairs could be phase-aligned, and the corresponding $d_{\text{orbit}}$ estimates should therefore be interpreted cautiously because of the small and potentially selected subset. The grey shaded region marks the long-transient-dominated window, $w_S = 57/180$–$63/180$, where reliable phase-aligned estimates are largely unavailable because only a small fraction of trajectory pairs converge to detectable cycles. Where sufficient aligned-cycle pairs are available, $d_{\text{orbit}}$ remains close to $d_{\text{final}}$, indicating that phase offsets within periodic dynamics account for only a small fraction of the measured trajectory separation. (b) Convergence outcomes for the same 500 trajectory pairs at each $w_S$. Stacked bars classify each pair into three mutually exclusive categories: both trajectories converged to detectable cycles, one trajectory converged while the other remained long-transient, or both trajectories remained long-transient throughout the 100,000-generation observation window. Hamming-distance measurements were recorded over the first 500 generations, whereas the convergence categories in panel (b) were determined using the full 100,000-generation observation window. Parameters: $S_{\text{low}} = 1/18$, $B_{\text{low}} = 6/18$, $B_{\text{high}} = 16/18$, $w_B = 10/18$, $\Delta_{\text{low}} = 5/18$, grid size $100 \times 75$, 500 paired random initial conditions per parameter setting, and initial density 0.25.

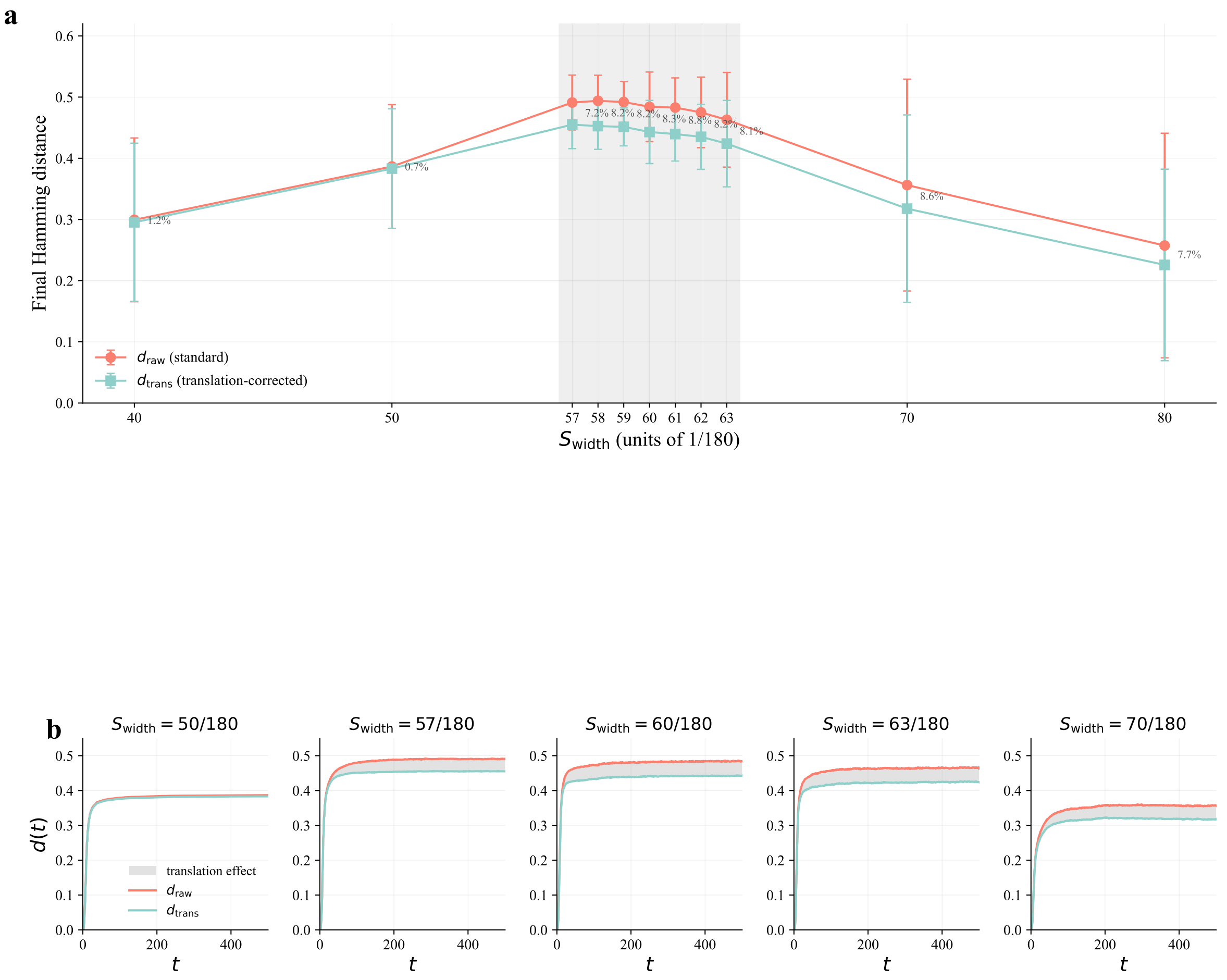

**Figure S11:** (a) Standard and translation-corrected final Hamming distance across $S_{width}$. The red circles show $d_{raw}$ — the conventional final Hamming distance $d_{final}$ averaged over 500 run pairs per $S_{width}$. The teal squares show $d_{trans}$ — the Hamming distance computed after optimally shifting one grid relative to the other via 2D FFT cross-correlation to minimise spatial translational offsets. Error bars indicate $\pm 1$ standard deviation. Numeric labels between the two curves indicate the per-run mean translation effect ratio $1 - d_{trans}/d_{raw}$ as a percentage. The grey shaded region highlights the long-transient-dominated zone ($S_{width} = 57/180 - 63/180$). The translation correction reduces the measured distance by 8% across the overlapping-window regime ($S_{width} \geq 57/180$) and by less than 2% in the fully separated regime ($S_{width} \leq 50/180$). Across the overlapping-interval regime, translation correction reduces the measured Hamming distance by approximately 8%, whereas the reduction is below 2% in the fully separated regime. Thus, global translational offsets account for only a minor component of the measured trajectory separation. Parameters: $S_{low} = 1/18, B_{low} = 6/18, B_{high} = 16/18, B_{width} = 10/18, \Delta_{low} = 5/18$, grid size 100 × 75, 500 generations per run, 500 random initial conditions per parameter set, initial density 0.25. (b) Time evolution of $d_{raw}(t)$ and $d_{trans}(t)$ for selected $S_{width}$ values. The Hamming distance $d(t)$ is shown for both the standard measurement (red) and its translation-corrected counterpart (teal), averaged over 500 run pairs per $S_{width}$. The grey shaded region between the two curves shows the reduction in measured Hamming distance produced by the optimal translation correction. Parameters: $S_{low} = 1/18, B_{low} = 6/18, B_{high} = 16/18, B_{width} = 10/18, \Delta_{low} = 5/18$, grid size $100 \times 75$, 500 generations per run, 500 random initial conditions per parameter set, initial density 0.25.

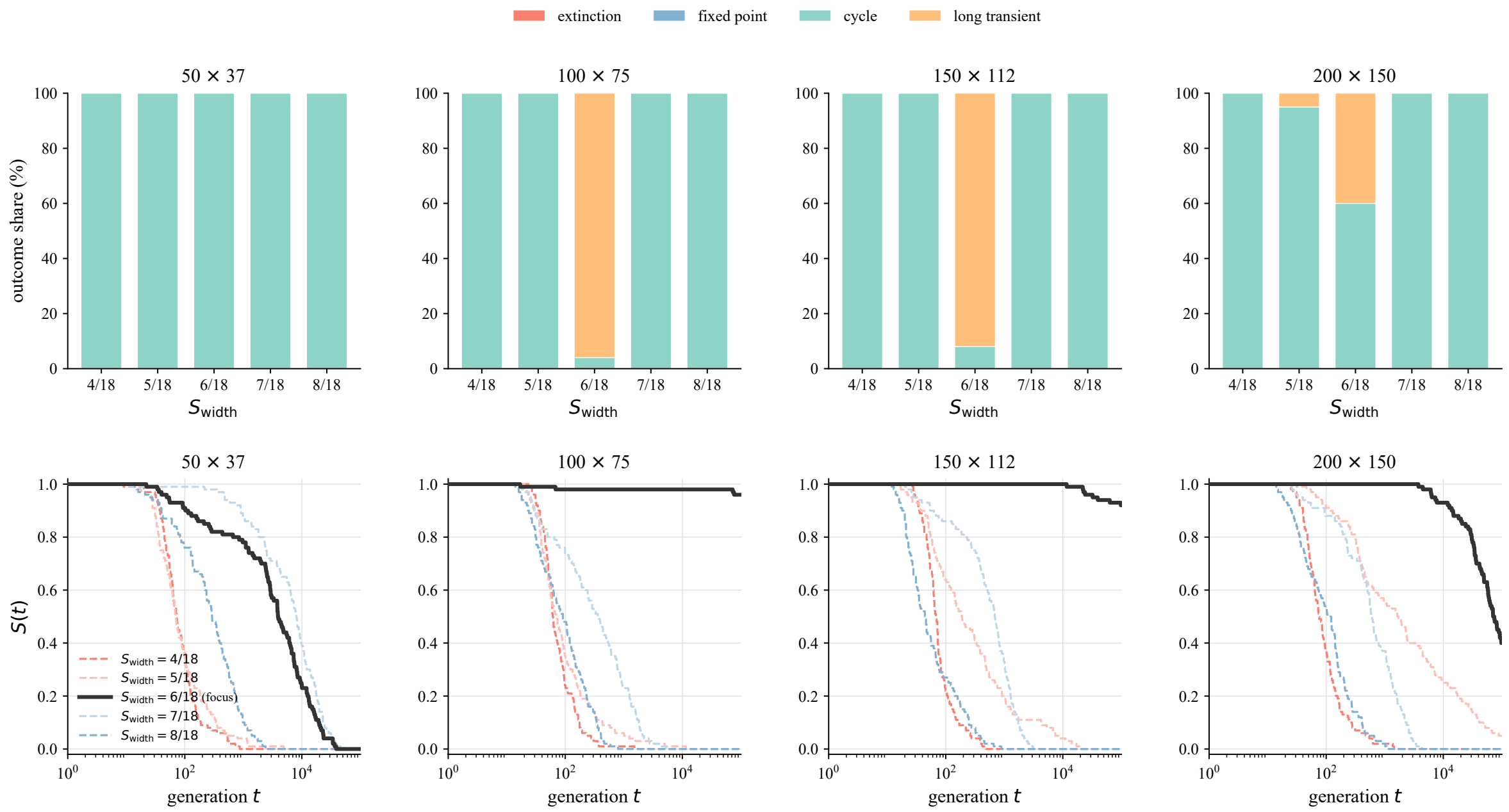

**Figure S12:** Finite-size dependence of outcome composition and transient lifetime. Top row: outcome shares for $w_S = 4/18$–$8/18$ on grids of size $50 \times 37$, $100 \times 75$, $150 \times 112$, and $200 \times 150$. Each bar summarizes 100 independent runs using a 100,000-generation observation window. Bottom row: Kaplan–Meier-style survival curves for the same five $w_S$ values and four grid sizes, showing the fraction of trajectories that have not yet reached extinction, a fixed point, or a detected periodic cycle by generation $t$. The $w_S = 6/18$ curve is highlighted as a reference value. Parameters: $S_{low} = 1/18, B_{low} = 6/18, B_{high} = 16/18, B_{width} = 10/18, \Delta_{low} = 5/18$, 100,000 generations per run, 100 random initial conditions per parameter set, initial density 0.25.

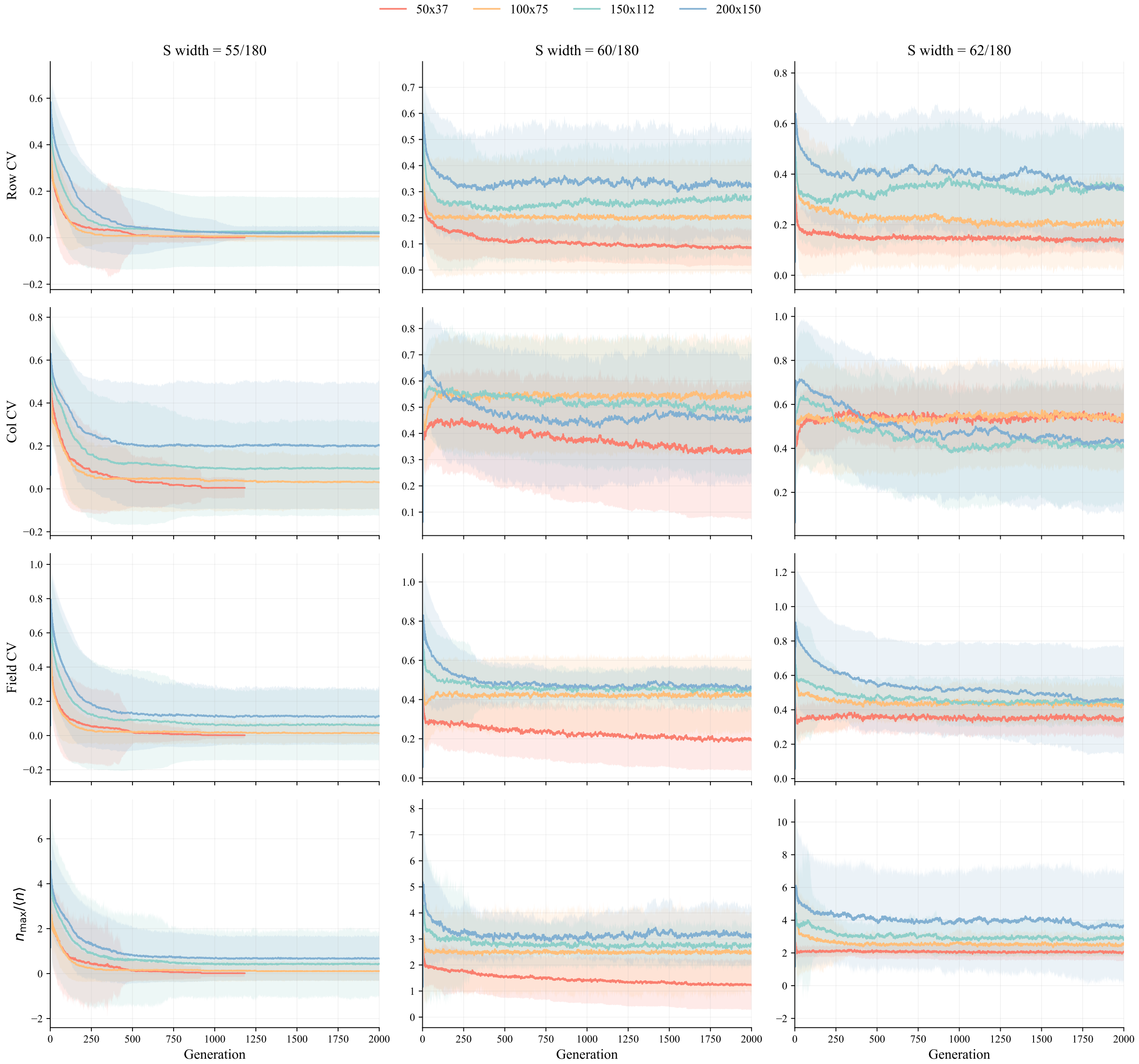

**Figure S13:** Evolution of field-organization statistics across grid sizes at three $w_S$ values. The four rows show, from top to bottom, the coefficient of variation (CV) of the row activity profile, the CV of the column activity profile, the CV of the field, and the ratio $n_{\max}/\langle n \rangle$, as functions of generation $t$. Parameters: $S_{low} = 1/18, B_{low} = 6/18, B_{high} = 16/18, B_{width} = 10/18, \Delta_{low} = 5/18$, 2000 generations per run, 100 random initial conditions per parameter set, initial density 0.25.